\documentclass[%
 reprint,
 amsmath,amssymb,
 aps,
]{revtex4-2}

\usepackage{graphicx}
\usepackage{dcolumn}
\usepackage{bm}
\usepackage{feynmp-auto}
\usepackage{slashed}
\usepackage{xcolor}
\usepackage{hyperref}
\usepackage{comment}
\usepackage{soul}

\begin{document}

\preprint{APS/123-QED}

\title{Axionic Acoustic Phonons from Weyl Semimetals}

\author{Joan Bernabeu}
\email{joan.bernabeu@uam.es}
\affiliation{Departamento de F\'isica de la Materia Condensada, Universidad Aut\'onoma de Madrid, Cantoblanco, E-28049 Madrid, Spain}
\author{Alberto Cortijo}%
\email{alberto.cortijo@csic.es}
\affiliation{Instituto de Ciencia de Materiales de Madrid (ICMM), Consejo Superior de Investigaciones Científicas (CSIC),\\
Sor Juana Inés de la Cruz 3, 28049, Madrid, Spain}%

\date{\today}

\begin{abstract}
The sound propagation properties resulting from a dynamical axion insulator generated from a Weyl semimetal are explored. Axial electron-phonon coupling in Weyl semimetals hybridize phonons with the axion field changing their velocity and attenuation. At the Random Phase Approximation (RPA) level, this hybridization gives rise to weaker modification of the speed of sound in longitudinal waves propagating parallel to the Axionic Charge Density Wave (ACDW) wavevector than in transverse waves.  On the other hand, the attenuation is modified by anharmonic phonon interactions beyond the RPA, but the axionic contribution to this attenuation is negligible. While the corrections on sound attenuation might be too weak to observe, the corrections to the speed of sound should be accessible in ultrasound propagation experiments. In the presence of a magnetic field, these corrections are sensitive to the magnetic field through the value of the ACDW gap and the axion velocity, but not through the axial anomaly, implying that these effects can be observed in the absence of an applied electric field. Finally, we also discuss the importance of axial electron-phonon interactions with respect to the more conventional vector interactions in these systems.
\end{abstract}

\maketitle

\section{Introduction}

Axions are the dynamical quanta of the axionic field, originally introduced long ago in high energy physics as a solution to the conundrum of the Strong CP probem in Quantum Chromodynamics \cite{peccei1977CP,peccei1977Constraints,wilczek1978Problem,weinberg1978ANew}. This fact, along with their robustness as a cold dark matter candidate \cite{preskill1982Cosmology,abbott1982acosmological,dine1982thenot,adams2022axion}, precipitated the proliferation of axion theories in the field. More recently, analogous systems were proposed in condensed matter \cite{li2010axions}. These are often referred to as dynamical axion insulators, so as to distinguish them from "normal" axion insulators \cite{mong2010antiferromagnetic,sekine2021axion}. In the latter, the $\theta$ in the axionic magnetoelectric term $\theta \mathbf{E}\cdot\mathbf{B}$ is static, while in the former it is a dynamical entity $\theta(t,\mathbf{x})$. This dependence on spacetime leads to an alteration of Maxwell's equations \cite{Wilczek87} known as "Axion Electrodynamics". 

Although originally proposed in topological magnetic insulators, it was later suggested that an dynamical axionic state of matter could emerge from an interacting phase in Weyl Semimetals \cite{wang2013chiral}. In the absence of interactions, these materials host Weyl nodes in their momentum-space band spectrum. As a consequence of the Nielsen-Ninomiya theorem \cite{NIELSEN198120,NIELSEN1981173}, these Weyl nodes come in pairs, each component denoted "left" and "right" by an associated label called chirality, borrowed from the high energy nomenclature. A low-energy description around these points has four fermionic operators $\psi_{\chi,\sigma}$ due to the pseudospin $\sigma = \pm$ and the effective chiral degrees of freedom, $\chi = L/R$. 

The latter result in an effective global continuous symmetry $U(1)_\textrm{A}$ known as axial/chiral symmetry, whereby the system is at the classical level invariant under phase rotations with opposite sign for each chirality, i.e. $\psi_{L/R,\sigma} \rightarrow e^{\pm i\alpha}\psi_{L/R,\sigma}$. These transformations are of comparable, but unmistakably different, form to those the standard $U(1)_\textrm{V}$ phase symmetry, $\psi_{L/R,\sigma} \rightarrow e^{i\alpha}\psi_{L/R,\sigma}$. In much the same way that $U(1)_\textrm{V}$ vector fields, most prominently the electromagnetic 4-vector potential $A_\mu$, couple to fermions, so can $U(1)_\textrm{A}$ axial vector fields $A_{5,\mu}$. It is important to note that the latter are not true gauge fields in the sense that the global $U(1)_\textrm{A}$ symmetry cannot be extended to a local one, as in the case of $U(1)_\textrm{V}$, without violating electric charge conservation. In Weyl semimetal systems, it has been proposed that such axial vector fields can arise from elastic deformations \cite{Zhou_2013,cortijo2015elastic,arjona2018elastic}.

Another difference is that the $U(1)_\textrm{A}$ symmetry can be broken by terms in the effective Hamiltonian that mix fermion chirality. This mixing between Weyl nodes can be brought about by electronic interactions \cite{wang2013chiral,roy2015magnetic,roy2017interacting}, which will also induce a corresponding Nambu-Goldstone mode, the axion. The axion in this case is the phason of the an Axionic Charge Density Wave (ACDW), as the nodal points are located at different points in the Brillouin zone, then the mixing term is non-local in momentum space and inconmensurate with the lattice position (hence the CDW character). Models that induce this behavior require either a strong coupling \cite{nambu1961dynamical1,nambu1961dynamical2,wang2013chiral} or a not-so-strong coupling combined with a catalyzing element such as a magnetic field \cite{gusynin1995dimensional,roy2015magnetic} or a compensated chiral density imbalance between the Weyl nodes involved \cite{braguta2019catalysis}.

Currently the difficulty lies in observing specific signatures from these systems. Axionic polaritons \cite{li2010axions}, the excitations resulting from the mixing of photons and axions in the presence of a magnetic field, are yet to be identified. The ACDW phase was reported in (TaSe$_4$)$_2$I \cite{gooth2019axionic,shi2021charge}, however that claim is disputed \cite{sinchenko2022does}. Other possible identifying signatures have been proposed \cite{mckay2021optical,Yu2021Dynamical,Curtis2023Finite,Bernabeu2024Chiral,Bernabeu2024Hysteresis,gao2024detecting,smith2024}. Many of these proposals invoke the effects of the chiral anomaly through the application of external electromagnetic fields \cite{ramananomaly2020}, to which the axion field most obviously couples. Here we suggest an alternative approach, which is to study axion effects on the acoustic phonon spectrum, experimentally accessible by sound propagation measurements. These have been proposed \cite{pikulin2016chiral,rinkel2019influence,zhang2020quantum, antebi2021anomaly, sukhachov2021anomalous} and applied \cite{schindler2020strong,lalibert2020field,Ehmcke2021propagation} to observe the effects of vector or axial gauge fields in the sound velocity and attenuation in Weyl/Dirac semimetals. These effects are different from the ones analyzed in this work, as fundamentally the Dirac fermions of an ACDW phase are gapped, which automatically means that the electron-phonon coupling cannot attenuate phonons with energies below the ACDW gap $\Delta$ at the mean-field level.

In this work we will analyze the combined effects of electron-phonon coupling in the form of axial elastic vector fields and axions on the velocities and attenuation coefficients of acoustic phonons. In section II we will present the phonon and fermion theory and introduce the effective phonon-axion action. In section III we will consider the effects that quadratic corrections on the speed of sound in the material. In section IV we will study how the anharmonic interactions contribute to the sound attenuation coeffecient. Both section III and IV will consider the absence and presence of a magnetic field.

\section{\label{sec:level1} Microscopic Theory}
\subsection{Harmonic Phonon Lagrangian}
We consider that phonons are described classically by the standard theory of elasticity, written in terms of continuous lattice displacements $\mathbf{u} \equiv (u_1,u_2,u_3)$, (the strain tensor is defined as the symmetrized derivative of these displacements, $u_{ij}=\frac{1}{2}(\partial_i u_j+\partial_j u_i)$),
\begin{gather}\label{free_phonon_action}
    S_ \textrm{ph,0} = \int d^4x \ \frac{\rho}{2}\left[ -u_i\partial_t^2 u_i + u_iK_{ij}u_j\right]
\end{gather}
where $\rho$ is the ionic density of the material and $K_{ij}$ is the elastic constant tensor. We will assume that the underlying lattice of our system is isotropic, so this tensor depends on two independent parameters, that can be related to the longitudinal and transverse phonon velocities, $c_\textrm{L}$ and $c_\textrm{T}$ respectively,
\begin{equation}
    K_{ij} = c_\textrm{T}^2(\delta_{ij}\partial_i^2-\partial_i\partial_j) + c_\textrm{L}^2\partial_i\partial_j.
\end{equation}
Imposing translation invariance one finds that the eigensystem of (\ref{free_phonon_action}) consists of three modes where two are transverse with dispersion $\omega_{\textrm{T}} = c_\textrm{T} |\mathbf{q}|$ and one is longitudinal with dispersion $\omega_\textrm{L} = c_\textrm{L} |\mathbf{q}|$, where $\mathbf{q}$ is the sound wavevector.

\subsection{Fermionic Sector and Interactions}
For the fermion sector, we consider an action in the presence of both a normal and axial vector fields, $A_\mu$ and $\tilde{A}_\mu$ respectively. The axion charge density wave in the presence of these fields reads
\begin{gather}\nonumber
    S_f = \int d^4 x\Psi^\dagger \gamma^0\left[i\gamma^\mu\left(\partial_\mu - ieA_\mu -i\gamma^5 \tilde{A}_{5,\mu} \right)\right. \\
    \label{CDW_action}
    \left. -\Delta e^{i\gamma^5 (\theta + 2b\cdot x) }\right]\Psi
\end{gather}
where we have taken the Fermi velocity to be $v_f = 1$ but will recover it whenever it is necessary, and $\theta$ is the axion field. This action in the absence of the CDW term proportional to $\Delta$ represents a minimal model for a time-reversal-breaking Weyl SM, where the $\gamma$ matrices are taken in the Weyl representation such that $\Psi = (\psi_{L,+},\psi_{L,-},\psi_{R,+},\psi_{R,-})$. In the following we assume that the 4-vector potential $A_\mu$ will be purely external magnetic source. It will be either null, or take the symmetric gauge form $A_\mu = (0, -\frac{e}{2}\mathbf{x}\times\mathbf{B})$ in the presence of a constant magnetic field $\mathbf{B}$ ($e$ is the electron charge). Later in section \ref{discussion} we will consider the effects of other potential sources of $A_\mu$.  In addition, $\tilde{A}_{5,\mu} \equiv  b_\mu + A_{5,\mu}^\textrm{ph}$ where $2b_\mu$ is the internodal separation,
\begin{equation}\label{axial_el_ph_coupling}
A_{5,\mu}^\textrm{ph} = (0, g u_{ij}\hat{b}_j),
\end{equation} 
is the axial vector field induced by phonons \cite{cortijo2015elastic,arjona2018elastic}, and $g$ is the the electron-phonon coupling (EPC) constant. In this expression, $\hat{b}_i$ are the components of the normalized internodal separation. For concreteness in our model, we will assume that the Weyl nodes are separated in the z axis, i.e.  $\hat{\mathbf{b}} = \hat{\mathbf{z}}$.

The ACDW action (\ref{CDW_action}) can be transformed via an axial rotation of the form $\Psi \rightarrow e^{i\left(\frac{\theta}{2} + b\cdot x\right)\gamma^5}\Psi$, into
\begin{gather}\label{fermionic_action}
    S_f = \int d^4 x\Psi^\dagger \gamma^0[i\gamma^\mu(\partial_\mu - ieA_\mu -i\gamma^5 A_{5,\mu}) -\Delta]\Psi. 
\end{gather}
The axial gauge field in this action is now comprised by the sum of separate phononic and axionic components, $A_{5,\mu} = \frac{1}{2}\partial_\mu\theta + A_{5,\mu}^\textrm{ph}$.
In principle we should consider as well the anomalous  contribution \cite{landsteiner2016notes}
\begin{equation}\label{anomaly_action}
    S_\textrm{an} = \frac{1}{8\pi^2}\int d^4x(2\mathbf{b}\cdot\mathbf{x} + \theta)\left(e^2\mathbf{E}\cdot\mathbf{B} + \frac{1}{3}\mathbf{E}_5\cdot\mathbf{B}_5\right),
\end{equation}
where the vector and axial electromagnetic fields are given by $\mathbf{E} = \partial_t \mathbf{A} - \nabla c_lA_0$, $\mathbf{B} = \nabla \times \mathbf{A}$, $\mathbf{E}_5 = \partial_t \mathbf{A}_5^\textrm{ph}$, and $\mathbf{B}_5 = \nabla \times \mathbf{A}_5^\textrm{ph}$ , where $c_l$ is the speed of light. Since we are not considering external electric fields, the term proportional to $\mathbf{E}\cdot\mathbf{B}$ will have no impact in the physics we are describing. As mentioned previously, other sources for $A_\mu$ will be considered in section \ref{discussion}. Concerning the term proportional to $\mathbf{E}_5\cdot\mathbf{B}_5$, there are two terms we need to take into account, one proportional to the internodal separation $\mathbf{b}$ and another to the axion field $\theta$. For now we will ignore their effects, and argue later in section \ref{discussion} why this is valid assumption.

By integrating out fermions and calculating the response to the axial gauge field, we are now able to derive a theory for the coupled acoustic phonon-axion system,
\begin{gather}\label{effective_action}
    S_\textrm{eff} - S_ \textrm{ph,0}= -i\textrm{Tr}\ln\left(G_0^{-1} + \gamma^\mu\gamma^5A_{5,\mu}\right)  = \sum_{n=2}^{\infty}S^{(n)}_\textrm{eff} \\
    \textrm{with} \quad S^{(n)}_\textrm{eff} \equiv -i\frac{(-1)^n}{n}\textrm{Tr}\left(G_0 \gamma^\mu\gamma^5A_{5,\mu}\right)^n,
\end{gather}
where $G_0^{-1} \equiv i\gamma^\mu\partial_\mu + e\gamma^\mu A_\mu - \Delta$ is the inverse fermionic  Green's function. In the last line we have omitted the term $\textrm{Tr}\left(G_0^{-1}\right)$ as it is independent of the bosonic fields, as well as the $n=1$ term in the infinite series as it vanishes.
One obvious yet important distinction between normal Weyl systems and the dynamical axion insulators is that the fermion system in the former is gapless ($\Delta=0$), whereas in the latter it is gapped ($\Delta\neq0$). This is important as response functions, i.e. the terms proportional to each power of $A_{5,\mu}$ appearing in the series in (\ref{effective_action}), in the gapped phase differ significantly from the gapless case.

\section{Harmonic Corrections}

\begin{figure}
    \centering
    \begin{fmffile}{bubble}
    \begin{fmfgraph*}(120,120) 
    \fmfleft{i} 
    \fmfright{o}
    
    \fmf{photon,tension=5, label=$q$}{i,v1} 
    \fmf{photon,tension=5,label=$q$}{v2,o}
    \fmf{plain,left=1, tension=1, label=$l$}{v1,v2}
    \fmf{plain,left=1, tension=1, label=$l-q$}{v2,v1}
    \fmfdot{v1,v2}
    \fmfv{label=$\gamma^\mu\gamma^5$, label.angle=0}{v1}
    \fmfv{label=$\gamma^\nu\gamma^5$, label.angle=180}{v2}

    \end{fmfgraph*}
    \end{fmffile}
    \caption{Diagrammatic representation of $\Pi^{\mu\nu}(q)$.}
    \label{fig:bubble_diagram}
\end{figure}
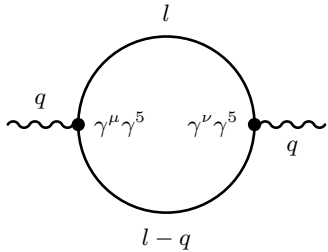

Here we are concerned with the quadratic corrections given by the term of order $n=2$ in (\ref{effective_action}). This term, indepedently of whether one considers the presence of a magnetic field, can be expressed in 4-momentum space, to the lowest order in 4-momentum $q$, as
\begin{gather}\label{action_quadratic_correction}
    S^{(2)}_\textrm{eff} = \int_q \frac{1}{2} \Pi^{\mu\nu}(q)[\mathcal{A}_{\mu a}U_a](-q)[\mathcal{A}_{\nu b}U_b](q)
\end{gather}
where the axial vector fields $A_{5,\mu} = \mathcal{A}_{\mu a} U_a$ have been expressed through a canonically normalized form of the constitutive displacement and axion fields, $U \equiv (\sqrt{\rho}\mathbf{u}, f_\theta \theta) \equiv (U_1, U_2, U_3, U_4)$. The constant $f_\theta \equiv \sqrt{\Pi^{00}(0)/4}$ is also referred to the axion decay constant in the context of high energy physics, and the matrix $\mathcal{A}$  is given by
\begin{gather}\label{A_matrix}
    \mathcal{A}(q) = 
    \begin{pmatrix}
        0 & 0 & 0 & (2f_\theta)^{-1}\omega \\
        \frac{\tilde{g}}{2}q_3 & 0 & \frac{\tilde{g}}{2}q_1 & (2f_\theta)^{-1}q_1 \\ 
        0 & \frac{\tilde{g}}{2}q_3 & \frac{\tilde{g}}{2}q_2 & (2f_\theta)^{-1}q_2 \\
        0 &  0 & \tilde{g} q_3 & (2f_\theta)^{-1}q_3
    \end{pmatrix}
\end{gather}
and where $\tilde{g} \equiv g/\sqrt{\rho}$. Finally, the axial polarization tensor $\Pi^{\mu\nu}$ can be derived from the bubble diagram in Fig. \ref{fig:bubble_diagram} . At lowest order in $q$, it has the form
\begin{gather}\label{main_text_polarization}
    \Pi^{\mu\nu}(0) = 4f_\theta^2
    \begin{pmatrix}
    1 & 0 & 0 & 0 \\
    0 & -v_1^2 & 0 & 0\\
    0 & 0 & -v_2^2 & 0\\
    0 & 0 & 0 & -v_3^2
    \end{pmatrix}
\end{gather}
where $v_i$ are the velocity components in the Cartesian direction $i$. In the model at hand, the velocities are possibly non-isotropic since we allow for the presence of a magnetic field.
Combining (\ref{action_quadratic_correction}) with $S_\textrm{ph.,0}$ in momentum space, the equations of motion for the bosonic fields can be derived. For arbitrary $\mathbf{q}$, one finds that the phononic modes propagating in the plane spanned by $\mathbf{b}$ and $\mathbf{q}$ mix with the axion. In the case where $\mathbf{q}$ is parallel or perpendicular to $\mathbf{b}$, only the mode parallel to $\mathbf{b}$, i.e. $U_3$, mixes with the axion. In the former case it is self-explanatory (the two transverse modes can always be chosen out of plane for the collinear $\mathbf{q}$ and $\mathbf{b}$), whereas the latter is due to the fact that when $\mathbf{q} \perp \mathbf{b}$, the harmonic corrections in (\ref{action_quadratic_correction}) only couple $U_3$ and $U_4$. For simplicity, we will now focus on these cases, with results based on the analysis in Appendix \ref{appendix_phonon_velocities}. The $(U_3,U_4)$ sector mixing is then described by the set of coupled equations
\begin{gather}\label{mix_matrix}
    \mathbf{q}^2\begin{pmatrix}
    c_3^2 + v_{\hat{\mathbf{q}}}^2r_3^2\xi^2 & v_{\hat{\mathbf{q}}}^2r_3\xi \\
    v_{\hat{\mathbf{q}}}^2r_3\xi  & v_3^2
    \end{pmatrix}
    \begin{pmatrix}
        U_3 \\
        U_4
    \end{pmatrix}
    = \omega^2
    \begin{pmatrix}
        U_3 \\
        U_4
    \end{pmatrix}
\end{gather}
where $c_3$ is the phonon velocity in the $\mathbf{b}$ direction, $v_{\hat{\mathbf{q}}}$ is the fermion velocity in the direction of $\mathbf{q}$ and we have defined the dimensionless parameters $\xi \equiv f_\theta\tilde{g}$ (dimensionless coupling constant), and $r_3=1$ if $c_3 = c_\textrm{T}$ and $r_3 = 2$ if $c_3 = c_\textrm{L}$. Considering the case where $\mathbf{q}\parallel\mathbf{b}$, this mixing results in corrections $\delta c_\textrm{L}$ to the longitudinal phonon velocity $c_\textrm{L}$. At the lowest order in $\xi$,
\begin{gather}\label{collinear_correction_max}
    (\delta c_\textrm{L})^2(\mathbf{q}\parallel\mathbf{b}) = - 4c_\textrm{L}^2\xi^2 + O(\xi^4).
\end{gather}
One can observe that instead of obtaining a corrrection to the phonon velocity proportional to $v_{\hat{\mathbf{q}}}^2$ in the absence of an axionic field, as (\ref{mix_matrix}) would indicate, the actual correction to the phonon velocity is proportional to $c_\textrm{L}^2$ and hence much more suppressed, since the Fermi velocity is typically of the order of two orders of magnitude greater than the phonon velocities (for instance in ACDW candidate (TaSe$_4$)$_2$I these are of the order of $ 10^3$ m/s \cite{saintpaul1988ultrasonic}). As mentioned previously, in the case $\mathbf{q}$ and $\mathbf{b}$ are perpendicular, $\mathbf{q}\perp\mathbf{b}$, the longitudinal mode is decoupled and $(\delta c_\textrm{L})^2(\mathbf{q}\perp\mathbf{b}) = 0$.  On the other hand, if $\mathbf{q} \parallel \mathbf{b}$, the two remaining transverse modes are decoupled from the axion yet they still receive corrections from their coupling to the fermion condensate,
\begin{gather}\label{transverse_corrections_max}
    (\delta c_{\textrm{T}i})^2(\mathbf{q}\parallel\mathbf{b}) = v_{i}^2 \xi^2 \quad \textrm{for } i=1,2.
\end{gather}
In the opposite scenario, when $\mathbf{q}\parallel \mathbf{x}\perp\mathbf{b}$, the transverse mode oscillating parallel to $\mathbf{q}$ receives the correction, $(\delta c_{\textrm{T}1})^2(\mathbf{q}\perp\mathbf{b}) = -c_\textrm{T}^2\xi^2$, analogously to (\ref{collinear_correction_max}), while the other transverse mode receives no corrections, $(\delta c_{\textrm{T}2})^2(\mathbf{q}\perp\mathbf{b}) = 0$.

In the absence of a magnetic field, i.e., when the axion propagates isotropically, $v_i = v_f$, the ratio between the ACDW corrections to the phonon velocities for longitudinal (Eq.(\ref{collinear_correction_max})) and transverse propagation (Eq.(\ref{transverse_corrections_max})) is independent of the EPC and yields,
\begin{equation}\label{quotients_no_B}
    \frac{\delta c_\textrm{L}(\mathbf{q}\parallel\mathbf{b})}{\delta c_\textrm{T}(\mathbf{q}\parallel\mathbf{b})} =-2\frac{c_\textrm{L}}{v_f}, \quad \textrm{if } \mathbf{B} = 0.
\end{equation}

\begin{figure}
    \begin{tabular}{cc}
      (a)   &  \\
       & \includegraphics[width = 0.95\linewidth]{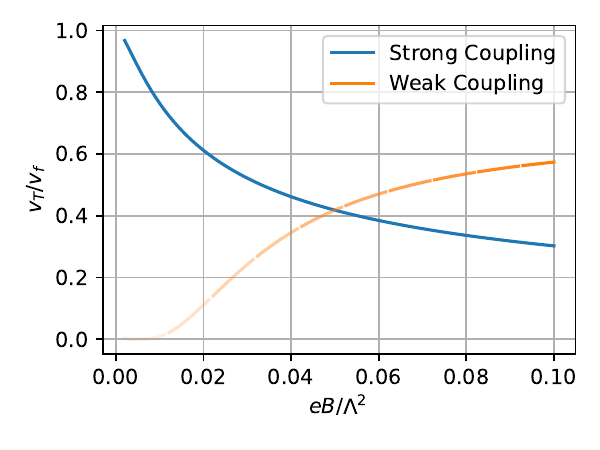} \\
       (b) & \\
       & \includegraphics[width = 0.21\textwidth]{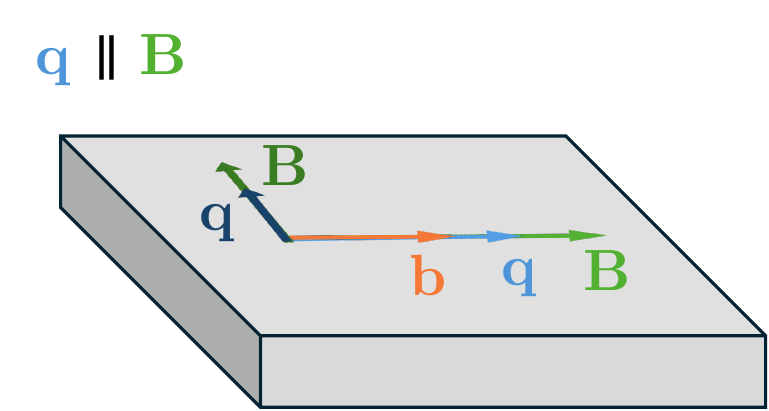} 
    \includegraphics[width = 0.21\textwidth]{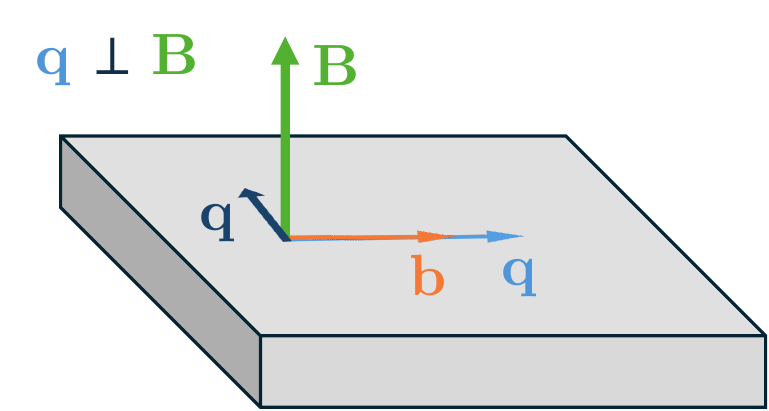} 
    \end{tabular}

    \caption{(a) Schematic representation of the axion transverse velocity $v_\textrm{T}$ in the strongly interacting (blue) and weakly interacting cases (orange) for varying magnetic field. $\Lambda$ is a UV cutoff provided by the effective theory. The vanishing of the weakly-interacting curve indicates that at low magnetic field, the condensate is exponentially small \cite{gusynin1995dimensional} and the effective axion action is not an appropriate description. (b) Experimental setup described in text. In the left image, the magnetic field $\mathbf{B}$ (green) and sound propagation wavevector $\mathbf{q}$ (blue) are parallel and are oriented parallel (light green) or perpendicular (dark green) to the  CDW wavector $2\mathbf{b}$ (orange). In the right image, the setup is the same but now $\mathbf{B}$ is perpendicular to the $\mathbf{q}-\mathbf{b}$ plane. Each case will yield either (\ref{quotients_B_parallel}) or (\ref{quotients_B_perpendicular}).}
    \label{fig:magnetic_field_stuff}
\end{figure}

When an external magnetic field is applied, the above scheme changes. While the propagation parallel to the magnetic field remains unaltered, i.e., the axion velocity along this direction is still $v_f$, along the transverse directions to $B$ will acquire a magnetic field-dependent velocity $v_\textrm{T} < v_f$.
In this scenario, to probe the effect of EPC and the axion mode, one can again distinguish the effects on the longitudinal and transverse phonon velocities. We thus consider the same geometry as before, where $\mathbf{q}$ and $\mathbf{b}$ are coplanar. In a first situation, one aligns $\mathbf{B}$ with $\mathbf{q}$, as shown in the left image of Figure \ref{fig:magnetic_field_stuff}(b). One finds that
\begin{equation}\label{quotients_B_parallel}
    \frac{\delta c_\textrm{L}(\mathbf{q}\parallel\mathbf{b})}{ \delta c_\textrm{T}(\mathbf{q}\parallel\mathbf{b})} = -2\frac{c_\textrm{L}}{v_\textrm{T}},\quad \textrm{if } \mathbf{q} \parallel \mathbf{B}.
\end{equation}
One can also consider the case where $\mathbf{B}$ is set perpendicular to the $(\mathbf{q}-\mathbf{b})$ plane, as in the right image of Figure \ref{fig:magnetic_field_stuff}(b). Now the two transverse modes no longer behave the same. Assuming that $\mathbf{B} \parallel \hat{\mathbf{x}}$,
\begin{equation}\label{quotients_B_perpendicular}
    \frac{\delta c_\textrm{L}(\mathbf{q}\parallel\mathbf{b})}{\delta c_\textrm{T,1}(\mathbf{q}\parallel\mathbf{b})} = -2\frac{c_\textrm{L}}{v_f}, \quad \frac{\delta c_\textrm{L}(\mathbf{q}\parallel\mathbf{b})}{\delta c_\textrm{T,2}(\mathbf{q}\parallel\mathbf{b})} = -2\frac{c_\textrm{L}}{v_\textrm{T}},\quad \textrm{if } \mathbf{q} \perp \mathbf{B}.
\end{equation}
Since $v_\textrm{T}$ is sensitive to $B$, the ratios in (\ref{quotients_B_parallel}-\ref{quotients_B_perpendicular}) should be $B$-dependent, further manifesting the coupling between the phonon and fermion sectors. The dependence on $B$ can be divided into two cases: One where the ACDW phase is driven by a strong coupling \cite{wang2013chiral} and the condensate $\Delta$ is approximately independent of the value of $B$, and one where the magnetic field has acted as a catalyzer  for the transition to the ACDW phase \cite{roy2015magnetic}, in which case the condensate does depend on $B$. The dependence of $v_\textrm{T}$ with the magnetic field in both cases is schematically displayed in Fig. \ref{fig:magnetic_field_stuff}(a).

\section{Anharmonic Corrections}
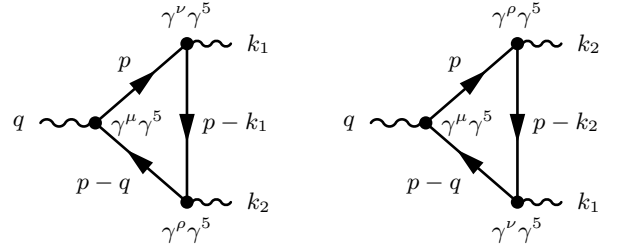
\begin{figure}
    \centering
    \vspace{0.5cm}
    \begin{tabular}{ccc}
    \begin{fmffile}{triangle}
\begin{fmfgraph*}(80,60)
    \fmfleft{i1,i2,i3}
    \fmfright{o1,o2,o3}
    
    \fmf{photon}{b1,o1}
    \fmf{photon}{i2,b2}
    \fmf{photon}{b3,o3}
    \fmf{phantom,tension=0.35}{b1,i1}
    \fmf{phantom,tension=0.35}{b2,o2}
    \fmf{phantom,tension=0.35}{b3,i3}
    \fmffreeze
    \fmf{plain_arrow, label=$p-k_1$, label.side=left}{b3,b1}
    \fmf{plain_arrow, label=$p-q$, label.side=left}{b1,b2}
    \fmf{plain_arrow, label=$p$, label.side=left}{b2,b3}

    \fmfv{label=$q$,label.angle=180}{i2}
    \fmfv{label=$k_2$,label.angle=0}{o1}
    \fmfv{label=$k_1$,label.angle=0}{o3}
    
    \fmfv{label=$\gamma^\rho\gamma^5$,label.angle=-90}{b1}
    \fmfv{label=$\gamma^\mu\gamma^5$,label.angle=0}{b2}
    \fmfv{label=$\gamma^\nu\gamma^5$,label.angle=90}{b3}

    \fmfdot{b1,b2,b3}
    \end{fmfgraph*}
    \end{fmffile}
    & \quad \quad \quad \quad &
    \begin{fmffile}{triangle2}
    \begin{fmfgraph*}(80,60)
    \fmfleft{i1,i2,i3}
    \fmfright{o1,o2,o3}
    
    \fmf{photon}{b1,o1}
    \fmf{photon}{i2,b2}
    \fmf{photon}{b3,o3}
    \fmf{phantom,tension=0.35}{b1,i1}
    \fmf{phantom,tension=0.35}{b2,o2}
    \fmf{phantom,tension=0.35}{b3,i3}
    \fmffreeze
    \fmf{plain_arrow, label=$p-k_2$, label.side=left}{b3,b1}
    \fmf{plain_arrow, label=$p-q$, label.side=left}{b1,b2}
    \fmf{plain_arrow, label=$p$, label.side=left}{b2,b3}

    \fmfv{label=$q$,label.angle=180}{i2}
    \fmfv{label=$k_1$,label.angle=0}{o1}
    \fmfv{label=$k_2$,label.angle=0}{o3}
    
    \fmfv{label=$\gamma^\nu\gamma^5$,label.angle=-90}{b1}
    \fmfv{label=$\gamma^\mu\gamma^5$,label.angle=0}{b2}
    \fmfv{label=$\gamma^\rho\gamma^5$,label.angle=90}{b3}

    \fmfdot{b1,b2,b3}
    \end{fmfgraph*}
    \end{fmffile}
        \\
    & \quad &
    \end{tabular}

    \caption{Triangle diagrams yielding the boson interaction vertex $\Gamma^{\mu\nu\rho}(q)$ (\ref{three_boson_vertex}).}
    \label{fig:triangle_diagrams}
\end{figure}

\begin{figure}
    \centering
    \begin{tabular}{ccc}
    
    \begin{fmffile}{bosonic_bubble}
    \begin{fmfgraph*}(80,80) 
    \fmfleft{i} 
    \fmfright{o}
    
    \fmf{photon,tension=5, label=$q$}{i,v1} 
    \fmf{photon,tension=5,label=$q$}{v2,o}
    \fmf{photon,left=1, tension=0.5, label=$l$}{v1,v2}
    \fmf{photon,left=1, tension=0.5, label=$l-q$}{v2,v1}
    \fmfdot{v1,v2}
    \fmfv{label=$\Gamma_{(3)}$, label.angle=0}{v1}
    \fmfv{label=$\Gamma_{(3)}$, label.angle=180}{v2}

    \end{fmfgraph*}
    \end{fmffile} &
    \begin{fmffile}{bosonic_tadpole}
    \begin{fmfgraph*}(60,60)
    \fmfstraight 
    \fmfleft{i0,i1,i2}
    \fmfright{o0,o1,o2}
    \fmf{photon, label=$q$}{i0,c0,o0}
    \fmf{phantom}{i1,c1,o1}
    \fmf{phantom}{i2,c2,o2}
    \fmffreeze
    \fmf{photon,tension=2, label=$0$}{c0,c1}
    \fmf{photon,left=1,tension=1, label=$l$}{c1,c2}
    \fmf{photon,left=1,tension=1}{c2,c1}
    \fmfv{label=$\Gamma_{(3)}$, label.angle=-90}{c0}
    \fmfv{label=$\Gamma_{(3)}$, label.angle=90}{c1}
    \end{fmfgraph*}
    \end{fmffile} &
    \begin{fmffile}{bosonic_loop}
    \begin{fmfgraph*}(80,80) 
    \fmfleft{i} 
    \fmfright{o}
    
    \fmf{photon,tension=1, label=$q$}{i,v1} 
    \fmf{photon,tension=1,label=$q$}{v1,o}
    \fmf{photon,right=1, tension=0.75, label=$l$}{v1,v1}
    \fmfdot{v1}
    \fmfv{label=$\Gamma_{(4)}$, label.angle=-90}{v1}

    \end{fmfgraph*}
    \end{fmffile} \vspace{5mm} \\
    (a) & (b) & (c)
    \end{tabular}
    
    \caption{Bosonic bubble (a), tadpole (b) and loop (c) diagrams contributing to $\Pi_\textrm{B}$ at one loop order. Only a) contributes to the imaginary part. Cubic interactions are denoted by $\Gamma_{(3)}$, corresponding to $\Gamma^{\mu\nu\rho}$ in (\ref{action_cubic_correction}-\ref{three_boson_vertex}),  and quartic ones by $\Gamma_{(4)}$.}
    \label{fig:bosonic_diagrams}
\end{figure}
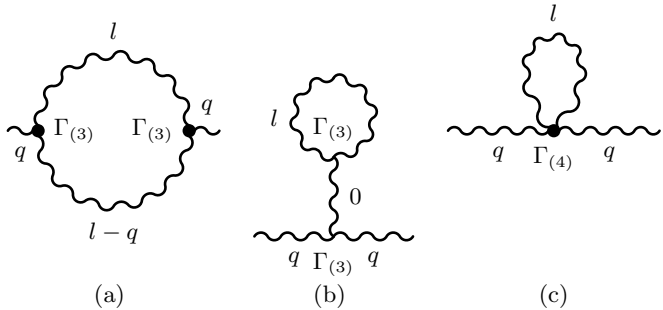

\begin{figure*}[t]
    \centering
    \begin{tabular}{c}
        \includegraphics[width = 0.99\textwidth]{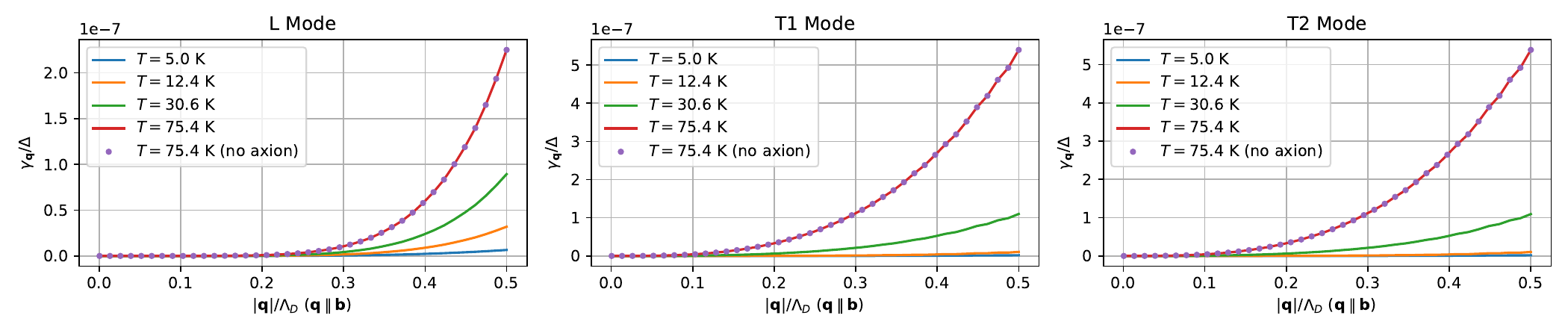} \\
        \includegraphics[width = 0.99\textwidth]{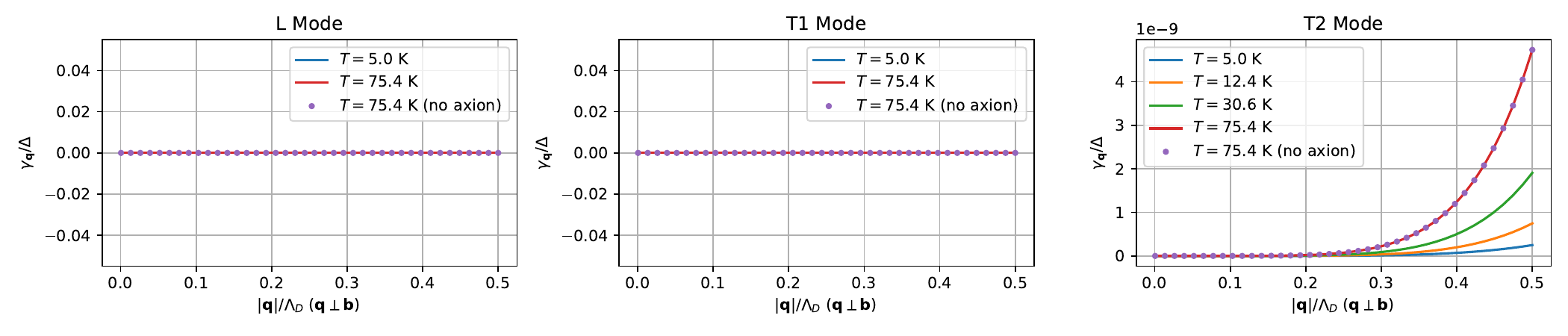}
    \end{tabular}

    \caption{Phonon linewidths as a function of wavevector modulus $q = |\mathbf{q}|$ for different temperatures $T$ in the parallel (top row) and perpendicular (bottom row) configurations. The wavevectors are referenced to a Debye momentum of $\Lambda_\textrm{D} = T_\textrm{D}/c_\textrm{L}$, where the Debye temperature is $T_\textrm{D} = 150 $K \cite{smontara2002anisotropy} and $c_\textrm{L} = 0.01 v_f = 3\cdot 10^4$ m/s. The linewidth is referenced with the ACDW gap of 0.23 eV as in (TaSe$_4$)$_2$I \cite{gooth2019axionic}. In the first row, the longitudinal mode linewidth scales with $\gamma_\mathbf{q}\propto |\mathbf{q}|^6$, whereas the transverse mode satisfies $\gamma_\mathbf{q}\propto |\mathbf{q}|^3$. In the bottom row, only the transverse mode oscillating parallel to $\mathbf{b}$ has a nonzero linewidth, which scales as $\gamma_\mathbf{q}\propto |\mathbf{q}|^6$. The case where the contribution by phonon-axion interactions is expressly disregarded is also shown for the temperature $T= 75.4 $ K, in dots. The curve traced by these dots follows the line that does include the axion-phonon interactions, which explictly illustrates the lack of impact of the axion mode $\theta$ in the phonon linewidths. Although not shown in the Figure, the axion also doesn't contribute to the phonon linewidths at other temperatures.}
    \label{fig:thermal_linewidths}.
\end{figure*}

\begin{figure}[t]
    \centering
    \includegraphics[width = 0.49\textwidth]{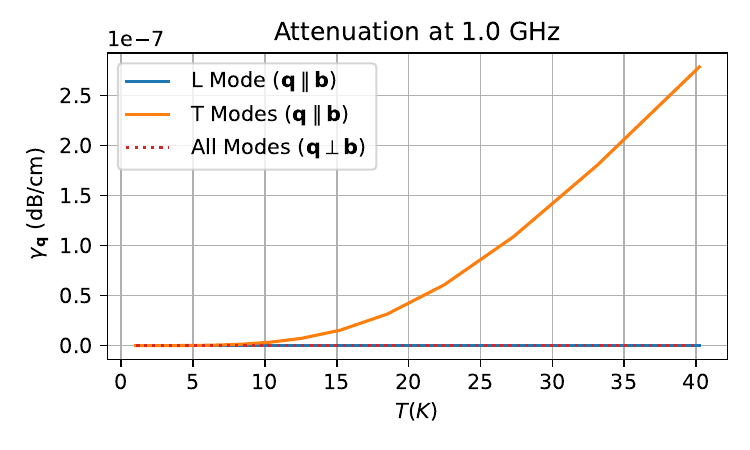}

    \caption{Sound attenuation for 1 GHz longitudinal and transverse sound modes at parallel and perpendicular relative orientations of $\mathbf{q}$ and $\mathbf{b}$. Attenuation for transverse modes in the $\mathbf{q} \parallel \mathbf{b}$ are many orders of magnitude stronger than all other modes. A Debye temperature of of 150 K is assumed \cite{smontara2002anisotropy}, while the energy cutoff for axions is taken to be the gap $\Delta = 0.23$ eV of (TaSe$_4$)$_2$I \cite{gooth2019axionic}. In addition, velocities of $v_f = 3\cdot10^6$ m/s and $c_\textrm{L} = 2c_\textrm{T} = 10^{-2}v_f$ are taken, as well as an EPC of $\bar{g}\Delta = g\Delta/\rho = 0.02$.}
    \label{fig:attenuation}
\end{figure}

Going beyond a quadratic expansion in the chiral vector field $A^\mu_5$, one can consider nonlinear effects, i.e., $S^{(n)}_\textrm{eff}$ for $n>2$. These correspond to anharmonic phonon-axion and phonon-phonon interactions. Here we truncate the expansion at $n=4$ so that only cubic and quartic interactions are considered, corresponding to $S^{(3)}_\textrm{eff}$ and $S^{(4)}_\textrm{eff}$ in (\ref{effective_action}). These interactions renormalize the phonon spectrum, leading not only to further corrections to the phonon velocities, but to finite phonon linewidths. In the case of the latter, one can show (see Appendix \ref{appendix_linewidths}) that the linewidth for the mode $a$ is given by
\begin{equation}\label{linewidth_formula}
    \gamma_{a,\mathbf{q}} = -\frac{\mathbf{u}^\dagger_{a,\mathbf{q}}\textrm{Im}\Pi_\textrm{B}(\omega_{a,\mathbf{q}},\mathbf{q})\mathbf{u}_{a,\mathbf{q}}}{2\omega_{a,\mathbf{q}}} 
\end{equation}
where where $\omega_{i,\mathbf{q}}$ and $\mathbf{u}_{i,\mathbf{q}}$ are the bosonic energies and polarizations of the system for the bosonic species $a$ in the absence of interactions between electrons and acoustic phonons, respectively, and $\Pi_\textrm{B}$ is the boson self-energy. Note that we are assuming boson propagators of the form $D_{0}^{-1} \sim (\omega^2 - \omega_\mathbf{q}^2)$ instead of the form $D_{0}^{-1} \sim (\omega^2 - \omega_\mathbf{q}^2)/(2\omega_\mathbf{q})$ which sometimes appears in the literature.

While at lowest order in perturbation theory, the contributing diagrams to $\Pi_\textrm{B}$ are the bubble, tadpole and loop diagrams (see Fig.\ref{fig:bosonic_diagrams}), only the bubble diagram has an imaginary component \cite{klein1967ultrasonic,PETHICK19661905,lang1999anharmonic,paulatto2015first} that results in a phonon linewidth. Since we will disregard further corrections to the real part of the spectrum which will be of a higher order in the EPC to those already calculated through the harmonic corrections, we will only require cubic interactions of $S^{(3)}_\textrm{eff}$ for the derivation of the imaginary part of the bubble diagram.

Formally, these cubic interactions read,
\begin{gather}\nonumber
    S^{(3)}_\textrm{eff} = \frac{1}{3} \int_{k_1,k_2}\Gamma^{\mu\nu\rho}(k_1,k_2) \cdot \\
    \label{action_cubic_correction}
    [\mathcal{A}_{\mu i}U_i](-k_1-k_2)[\mathcal{A}_{\nu j}U_j](k_1)[\mathcal{A}_{\rho k}U_k](k_2)
\end{gather}
where $\Gamma^{\mu\nu\rho}(k_1,k_2)$ needs to be calculated from the sum of the triangle diagrams in Fig. \ref{fig:triangle_diagrams}. Note that in the absence of a gap $\Delta$, these diagrams yield the axial anomaly \cite{adler1969axial,bell1969PCAC}. It should be unsurprising then that a certain momentum routing must be taken in order to preserve charge conservation. For a discussion on this calculation, see Appendix \ref{appendix_triangle_diagrams}. At lowest order in the external momenta, the vertex reads
\begin{gather}\label{three_boson_vertex}
    \Gamma^{\mu\nu\rho}(k_1,k_2) =  \frac{i}{3\pi^2}\epsilon^{\mu\nu\rho\sigma}(k_2-k_1)_\sigma +O(k_1^2,k_2^2).
\end{gather}
Note that the vertex at this order in external momenta is independent of the gap $\Delta$. Inserting (\ref{three_boson_vertex})  in (\ref{action_cubic_correction}), the imaginary part of the bubble diagram of Fig. \ref{fig:bosonic_diagrams}a, and hence the phonon linewidths (\ref{linewidth_formula}), can be obtained. We have to keep in mind that the integral over loop momentum is divergent, and a cutoff must be imposed. For phonons this should be a scale $\Lambda_\textrm{D}\sim T_\textrm{D}/c_\textrm{L}$, where $T_\textrm{D}$ is the Debye temperature. For the axion, this scale should be given by the cutoff of the effective axion theory, $\Delta/v_f$.

The resulting phonon linewidths are displayed in Fig.\ref{fig:thermal_linewidths} for the cases where phonons with momentum $\mathbf{q}$ propagate parallel (top row) and transverse (bottom row) to the Weyl node separation $\mathbf{b}$. It is found that, in both cases, the phonon mode polarized parallel to $\mathbf{b}$ follows a $\gamma_\mathbf{q} \propto |\mathbf{q}|^6$ scaling. On the other hand, the linewidth of modes polarized perpendicularly to $\mathbf{b}$ scales like $\gamma_\mathbf{q} \propto |\mathbf{q}|^3$ in the case where $\mathbf{q}$ is parallel to $\mathbf{b}$, whereas $\gamma_\mathbf{q} = 0$ for these modes when $\mathbf{q}$ is perpendicular to $\mathbf{b}$. The latter fact is due to the $\mathcal{A}_{\mu a}(q)$ factors are identically 0 for these phonon modes (\ref{A_matrix}). In addition, it can be checked that the axion does not play an important role in determining the phonon linewidth, as is displayed for the $T=75.4$ K case in Fig.\ref{fig:thermal_linewidths}.

As was the case for the harmonic corrections, the anharmonic corrections will also be sensitive to the effects of a magnetic field. For simplicity, we will discuss only the effects in the quantum limit, where only the lowest Landau level (LLL) is relevant. In this limit, the integrals over loop momentum components transverse to the magnetic field in the diagrams in Fig. \ref{fig:triangle_diagrams} lead to an $eB$ prefactor coming from the degeneracy of the LLL, as it happens in the case of the bubble diagram in Fig. \ref{fig:bubble_diagram} (see Appendix section \ref{appendix:bubble_finite_B} for the calculation of the bubble diagram at finite magnetic field). Since the contributions of these diagrams must have an adimensional prefactor before the external energy-momentum contribution (see e.g. Eq.(\ref{three_boson_vertex})), this diagram must be proportional to $eB/\Delta^2$. The remaining magnetic field dependence in the the cubic bosonic interaction in Eq.(\ref{action_cubic_correction}) could come from the axion decay constant $f_\theta$. However, as was the case with the zero-field case, we do not expect the axion to contribute to the phonon attenuation due to the reduced available phase space.

Therefore, since the phonon attenuation $\gamma_\mathbf{q}$ contains two vertex insertions $\Gamma^{\mu\nu\rho}$ (symbolically denoted as $\Gamma_{(3)}$ in Fig. \ref{fig:bosonic_diagrams}a ), then the dependence of the attenuation on the magnetic field comes from a leading factor of $(eB)^2/\Delta^4$. In this limit, therefore, the attenuation will be measured to be $\gamma_\mathbf{q}\propto B^2$ if the ACDW phase is generated primarily from interactions. However, if the ACDW phase transition is catalyzed by a magnetic field, then $\Delta$ will depend on $B$ and the overall prefactor will scale as $\gamma_\mathbf{q}\propto(eB)^2/\Delta^4 \propto e^{\frac{C}{eB}}$, where $C$ is some system-specific constant \cite{gusynin1995dimensional}.

\section{Discussion}\label{discussion}
\subsection{Effects of other electron-phonon interactions}
In the preceding sections we have focused solely on the axial interaction Eq.(\ref{axial_el_ph_coupling}) between electrons and acoustic phonons, disregarding the effects of interactions that are chirality-independent, such as the standard deformation potential coupling. In principle, corrections associated to this coupling can appear either in the chirally-rotated fermionic action (\ref{fermionic_action}) or from the anomaly (\ref{anomaly_action}). In the former case, it can be shown that the contributions of the axial interaction dominate vector interactions. In contrast to the axial couplings, the fermion bubble with vector couplings (which is the same as the one displayed in Fig.\ref{fig:bubble_diagram} but without the $\gamma^5$ in the vertices) is zero in energy-momentum at order $O(q^0)$ due to charge conservation (see Appendix \ref{appendix_bubble_diagrams}). This is in contrast to the axial bubble of Fig.\ref{fig:bubble_diagram}, which is non-zero at this same order. This is a consequence of imposing charge conservation ($U(1)_\textrm{V}$ gauge invariance) in the regularization scheme of the diagram in Fig. \ref{fig:bubble_diagram}. A similar result is found for the triangle diagrams analogous to Fig.\ref{fig:triangle_diagrams}. Imposing charge conservation on any variation of these diagrams that removes at least one of the $\gamma^5$ in the vertices leads to them vanishing at linear order in external momenta, as is shown in Appendix \ref{appendix_triangle_diagrams}. Again, this is opposed to the contribution of the diagrams in Fig.\ref{fig:triangle_diagrams}, that remain finite at linear order in $k_1$ and $k_2$ and result in Eq.(\ref{three_boson_vertex}).

Therefore, we are left to consider the effects of the anomaly (\ref{anomaly_action}). Concerning the vector component $\mathbf{E}\cdot\mathbf{B}$, the standard coupling between electrons and acoustic phonons is the deformation potential, which enters the action (\ref{fermionic_action}) as a scalar potential $A_0^\textrm{ph} = g'\textrm{Tr}(u_{ij})$, where $g'$ is a coupling constant. This by itself would not trigger the anomaly (\ref{anomaly_action}), as it does not generate an elastic magnetic field. However, in the presence of an external magnetic field, it does lead to a harmonic term of the form $\frac{g'eB}{\sqrt{\rho} f_\theta}(\mathbf{q}\cdot\mathbf{U})(\mathbf{q}\cdot\hat{\mathbf{B}})$ ($\hat{\mathbf{B}}$ is the unit vector in the direction of the magnetic field). In the case where $\mathbf{B}$ lies perpendicular to the plane spanned by $\mathbf{b}$ and $\mathbf{q}$ and therefore the coupling is zero, the ratio in Eq.(\ref{quotients_B_perpendicular}) remains unchanged. On the other hand, the relation (\ref{collinear_correction_max}) is modified for $\mathbf{q} \parallel \mathbf{B}$, and in principle $\delta c_3^2 \propto v_3^2$ so that the main effect of axion hybridization is lost. However, this is only the case when $\frac{eB}{f_\theta^2}\frac{g'}{g} \gtrsim 1$, so, for sufficiently weak magnetic fields, this effect is suppressed and the axion coupling mediated by (\ref{axial_el_ph_coupling}) will dominate. On the other hand, the axial component $\mathbf{E}_5\cdot\mathbf{B}_5$ in Eq.(\ref{anomaly_action}) contains two elements: One proportional to $\theta$, and another proportional to $\mathbf{x}\cdot\mathbf{b}$. The former term is just another axion-phonon coupling included in the expression (\ref{action_cubic_correction}). However, as was mentioned in the previous section and is displayed in Fig. \ref{fig:thermal_linewidths}, the phonon linewidth is not sensitive to the presence of the axion mode, so this term does not alter our results. Furthermore, while the coupling to $(\mathbf{b}\cdot\mathbf{x})\mathbf{E}_5\cdot\mathbf{B}_5$ might alter the phonon spectrum, the corrections are of cubic in energy-momentum, which goes beyond our quadratic expansion.

\subsection{Ultrasonic probes}
Considering the dynamical axion insulator, we have shown in Eq.(\ref{collinear_correction_max}) how the axion decreases the correction $\delta c_3$ to be proportional to the free phonon velocity $c_3$ instead of the electron velocity that would be obtained simply with a gap in the Dirac spectrum. On the other hand, while the axial fermion-phonon coupling (\ref{axial_el_ph_coupling}) does induce phonon decay through the anomalous vertex (\ref{three_boson_vertex}), the axion itself does not contribute to the linewidth as shown in Fig. \ref{fig:attenuation}. This is due to both suppression of the phase space available to the generation of virtual axions in the bosonic bubble in Fig. \ref{fig:bosonic_diagrams}a. and to the fact that their contribution is zero when both virtual bosons in the bubble are axions due to the antisymmetric form of (\ref{three_boson_vertex}).

A probe of these effects can be experimentally achieved by comparing the longitudinal and transverse sound velocities in Weyl semimetals close to the metal-ACDW phase transition, above and below the transition temperature $T_\textrm{CDW}$. These mesurements have been employed to study the imprint of the superconducting phase transition in Sr$_2$RuO$_4$ in the speed of sound \cite{Okuda2002Unconventional,Benhabib2021Ultrasound,ghosh2021thermodynamic}. In the Weyl SM phase 
the sound velocities are $c_\textrm{L}$ for longitudinal and $c_\textrm{T}$ for transverse propagation and the change from these velocities to those in the ACDW phase is precisely what is given in Eqs.(\ref{collinear_correction_max}-\ref{quotients_B_perpendicular}). The sensitivity of the velocities to the relative angle between $\mathbf{b}$, the phonon wavevector $\mathbf{q}$, and potentially magnetic field $\mathbf{B}$, can be exploited. This sort of anisotropy in sound propagation has already been studied considering the relative angle between the latter two vectors using an ultrasonic interferometer to study the imprint of Landau Levels on the phonon dispersion in the Weyl SM TaAs \cite{rinkel2019influence,lalibert2020field}. In our case the Weyl node separation $2\mathbf{b}$ is also an important ingredient when probing direction-dependent sound propagation due to its special role in the EPC (\ref{axial_el_ph_coupling}). Importantly, in the scenario presented here, these probes do not invoke the presence of an external electric field as is usually the case for probes of the axion which involve the response of the axial anomaly\cite{ramananomaly2020}. If a magnetic field is used, its imprint can affect the the results that depend on the axion velocity tranverse to the magnetic field, $v_\textrm{T}$, i.e., Eqs. (\ref{quotients_B_parallel}) and (\ref{quotients_B_perpendicular}).

Sound attenuation can also measured in the suggested ultrasound experiments. Comparing measurements between the Weyl SM phase and the ACDW phase might be more complicated, however, than for the speed of sound. In the former phase, the leading contribution to the phonon linewidth is derived from the imaginary part of $\Pi^{\mu\nu}(q)$, which is nonzero as the Weyl SM phase is gapless. It results in a phonon linewidth that is quadratic in the EPC. In the ACDW phase, however, the imaginary part of $\Pi^{\mu\nu}(q) = 0 $ for phonon frequencies below de CDW gap $\Delta$, and effects beyond RPA must be considered to obtain the effective anharmonic interactions[\ref{three_boson_vertex}]. The resulting contibution to the phonon linewidth (\ref{linewidth_formula}) is more suppressed than its metallic counterpart as it is proportional to the sixth power of the EPC. Therefore, a large drop-off  in $\gamma_\mathbf{q}$ is expected across the CDW transition, although measuring the effects below the transition is more complicated due to the aforementioned suppression by the EPC. Fig. \ref{fig:attenuation} shows attenuation for sound waves with frequencies around 1 GHz, for both cases $\mathbf{q} \parallel \mathbf{b}$ and $\mathbf{q} \perp \mathbf{b}$. Transverse modes in the former case have the highest attenuation. Due to the aforementioned suppression however, it is of the order of $10^{-9}-10^{-7}$ dB/cm in the $T=2-40$ K range.

\section{Closing Remarks}
Initial proposals for condensed matter axion detection involved mixing with other bosonic modes, particularly plasmons \cite{li2010axions}. More recent proposals consider the possibility that optical phonons themselves play the role of the dynamical axion in some symmetry-allowed Dirac insulators\cite{lhachemi2024phononic}. Similarly but also in contrast with these two possibilities, here we show that dynamical (fermionic interaction-induced) axions mix with the acoustic phonon modes. Starting from a minimal model for a Weyl SM-ACDW (\ref{fermionic_action}) coupled to acoustic phonons (\ref{free_phonon_action}) via an axial elastic vector field (\ref{axial_el_ph_coupling}), we have analyzed how the real and imaginary part of the acoustic phonon is modified. Since the fermion spectrum is gapped in the ACDW phase, the RPA only leads to modification of the sound velocity. Most notably, the axion inhibits stronger velocity corrections to longitudinal mode when  $\mathbf{q}\parallel\mathbf{b}$ (\ref{collinear_correction_max}) which can be tested through ultrasound experiments. Considering corrections beyond the RPA stemming from anomalous triangle diagrams (Fig. \ref{fig:triangle_diagrams}) lead to modifications in the attenuation spectrum, that are stronger in the transverse directions but nevertheless suppressed by the EPC. While this signal might be too weak to be detected through ultrasound probes, higher resolution could possibly be achieved through inelastic neutron scattering experiments (see, e.g., \cite{shapiro1972critical,pintschovius2005electron,pang2013phonon}). 

\begin{acknowledgments}
J.B. is supported by FPU grant FPU20/01087. A.C. acknowledges financial support from
the Ministerio de Ciencia e Innovación through the grant  PID2021-127240NB-I00. J.B. thanks Stanlisaw Galeski for insightful discussions.

\end{acknowledgments}

\appendix
\begin{widetext}

\section{Phonon Velocities}\label{appendix_phonon_velocities}
The quadratic Hamiltonian obtained from the sum of (\ref{free_phonon_action}) and (\ref{action_quadratic_correction}), in energy-momentum space, reads
\begin{gather}
    H =  \int_{\mathbf{q}}U_i(-q)\mathcal{H}(\mathbf{q})U_j(q) = \frac{1}{2}\int_{\mathbf{q}}U_i(-q)[c_\textrm{T}^2(\delta_{ij}\mathbf{q}^2-q_iq_j) + c_\textrm{L}^2q_iq_j + \Pi^{\mu\nu}\mathcal{A}_{\mu i}(0,-\mathbf{q})\mathcal{A}_{\nu j}(0,\mathbf{q})]U_j(q)
\end{gather}
where the zero in $\mathcal{A}_{\mu i}(0,-\mathbf{q})$ implies it is evaluated at zero energy. Taking $q_1 = 0$ without loss of generality, one finds that the matrix $H(\mathbf{q})$ takes the form
\begin{gather}\label{appendix_full_hamiltonian}
    \mathcal{H}(\mathbf{q}) = \begin{pmatrix}
        c_\textrm{T}^2\mathbf{q}^2 + v_1^2\xi^2q_3^2 & 0 & 0 & 0 \\
        0 & c_\textrm{L}^2q_2^2 + \left(c_\textrm{T}^2 + v_2^2\xi^2\right)q_3^2 & \left(c_\textrm{L}^2 - c_\textrm{T}^2+ v_2^2\xi^2\right)q_2 q_3 & v_2^2\xi q_2q_3\\
        0 & \left(c_\textrm{L}^2 - c_\textrm{T}^2+ v_2^2\xi^2\right)q_2q_3 &  \left(c_\textrm{T}^2+ v_2^2\xi^2\right)q_2^2 + \left(c_\textrm{L}^2 + v_3^2 4\xi^2\right)q_3^2 & v_2^2\xi q_2^2+  v_3^2 2\xi q_3^2 \\
        0 & v_2^2\xi q_2q_3 & v_2^2\xi q_2^2+  v_3^2 2\xi q_3^2 & v_2^2q_2^2 + v_3^2 q_3^2
    \end{pmatrix}
\end{gather}
It is clear how the mode $U_1$, perpendicular to $\mathbf{b}-\mathbf{q}$ plane is left uncoupled, as well as that when $\mathbf{q}$ and $\mathbf{b}$ are parallel or perpendicular, $U_2$ decouples too. Considering first the former case where $\mathbf{q}\parallel \mathbf{b}$, one finds
\begin{gather}\label{appendix_parallel_hamiltonian}
    \mathcal{H}(\mathbf{q}\parallel\mathbf{b}) = \begin{pmatrix}
        \left(c_\textrm{T}^2 + v_1^2\xi^2\right)\mathbf{q}^2 & 0 & 0 & 0 \\
        0 & \left(c_\textrm{T}^2 + v_2^2\xi^2\right)\mathbf{q}^2 & 0 & 0\\
        0 & 0 &    \left(c_\textrm{L}^2 + v_3^2 4\xi^2\right)\mathbf{q}^2 &  v_3^2 2\xi \mathbf{q}^2 \\
        0 & 0 &  v_3^2 2\xi \mathbf{q}^2 &  v_3^2 \mathbf{q}^2
    \end{pmatrix}
\end{gather}
It is immediately seen how the corrections to the transverse modes are those given by (\ref{transverse_corrections_max}). On the other hand, the velocities of the mixed longitudinal-axion modes are
\begin{gather}\label{appendix_full_velocity_correction}
    c_{\pm}^2 = \frac{c_\textrm{L}^2 + v_3^2(1+4\xi^2)}{2} \pm \frac{\sqrt{(v_3^2-c_\textrm{L}^2)^2 + 2v_3^2(v_3^2+c_\textrm{L}^2)4\xi^2 + v_3^4 16\xi^4}}{2}.
\end{gather}
By taking the leading order expansion in $\xi$, one recovers (\ref{collinear_correction_max}). On the opposite limit, when $\mathbf{b}\perp\mathbf{q}$, the Hamiltonian matrix reads
\begin{gather}\label{appendix_perpendicular_hamiltonian}
    \mathcal{H}(\mathbf{q} \perp \mathbf{b}) = \begin{pmatrix}
        c_\textrm{T}^2\mathbf{q}^2 & 0 & 0 & 0 \\
        0 & c_\textrm{L}^2\mathbf{q}^2  & 0 & 0\\
        0 & 0 &  \left(c_\textrm{T}^2+ v_2^2 \xi^2\right)\mathbf{q}^2 & v_2^2 \xi \mathbf{q}^2  \\
        0 & 0 & v_2^2\xi \mathbf{q}^2 & v_2^2\mathbf{q}^2
    \end{pmatrix}
\end{gather}
Now the uncoupled transverse mode as well as the longitudinal mode do not get any corrections from the electron-phonon coupling. The remaining mixed $U_3$ and axion modes read as (\ref{appendix_full_velocity_correction}) but with $c_\textrm{L}\rightarrow c_\textrm{T}$, $v_3\rightarrow v_2$ and $\xi\rightarrow\xi/2$. In the perturbative $\xi$ limit, this correction will thus be of the order of $c_\textrm{T}^2$, as is stated in (\ref{collinear_correction_max}). Since $v_i \gg c_\textrm{T}$, this correction to the transverse velocity is much smaller than the one obtained in the $\mathbf{q}\parallel \mathbf{b}$ case (\ref{appendix_parallel_hamiltonian}). Similarly, one also sees that the maximum correction to the longitudinal velocity is also given in the $\mathbf{q}\parallel \mathbf{b}$ case (\ref{appendix_parallel_hamiltonian}). This is why we concentrate on the the case $\mathbf{q}\parallel\mathbf{b}$ in (\ref{quotients_no_B}-\ref{quotients_B_perpendicular}).

An interesting aspect to notice is that a non-perturbative normalized electron-phonon coupling $\xi \propto \bar{g}f_\theta$ would lead to unconventional behavior such as phonon velocities of the scale of the Fermi velocity or behavior where the transverse phonons could propagate faster than longitudinal ones. Since this behavior is not empirically accurate, we do not consider it.
\section{Phonon Linewidths}\label{appendix_linewidths}
Assuming that the perturbation theory eigenstates have energies of the form $\omega_\mathbf{q} + i\gamma_\mathbf{q}$ where $\gamma_\mathbf{q}\ll \omega_\mathbf{q}$, the eigenenergies are found from solving the eigenvalue/eigenstate system
\begin{gather}
    [\omega_{\mathbf{q}}^2\mathbf{1} - M_0(\mathbf{q}) +\Pi_\textrm{F}(\mathbf{q})+ \Pi_\textrm{B}(\omega,\mathbf{q})]\mathbf{u}_{\mathbf{q}} = 0
\end{gather}
The F,B subscripts denote the fermionic and bosonic bubbles, i.e. those represented in Figures \ref{fig:bubble_diagram} and \ref{fig:bosonic_diagrams}a, respectively.. Using perturbation theory $\omega_{\mathbf{q}}^2 = \omega_{0,\mathbf{q}}^2 + \epsilon$, where
\begin{gather}
    [\omega_{0,\mathbf{q}}^2\mathbf{1} - M_0(\mathbf{q})]\mathbf{u}_{0,\mathbf{q}} = 0
\end{gather}
and $\epsilon$ is the perturbation generated by $\Pi_\textrm{F}$ and $\Pi_\textrm{B}$, one finds that
\begin{gather}
    \epsilon = -\mathbf{u}^\dagger_{0,\mathbf{q}}[\Pi_\textrm{F}(\omega_{0,\mathbf{q}},\mathbf{q}) + \Pi_\textrm{B}(\omega_{0,\mathbf{q}},\mathbf{q})]\mathbf{u}_{0,\mathbf{q}}.
\end{gather}
Since $\Pi_\textrm{F}(\mathbf{q})$ (\ref{main_text_polarization}) at lowest order in $q$ and $\mathbf{u}_{0,\mathbf{q}}$ are real ($M_0(\mathbf{q})$ is a symmetric matrix), one finds that
\begin{equation}
    \gamma_{\mathbf{q}} = \frac{\textrm{Im}\ \epsilon}{2\omega_{0,\mathbf{q}}} = - \frac{\mathbf{u}^\dagger_{0,\mathbf{q}}\textrm{Im}\Pi_\textrm{B}(\omega_{0,\mathbf{q}},\mathbf{q})\mathbf{u}_{0,\mathbf{q}}}{2\omega_{0,\mathbf{q}}} .
\end{equation}

\section{Fermionic Bubble Diagrams}\label{appendix_bubble_diagrams}

\subsection{Zero Magnetic Field}
We first consider the polarization tensor, which corresponds to the the bubble diagram in Figure \ref{fig:bubble_diagram} without the $\gamma^5$ matrix in the vertices. To do so, we employ the free fermion propagator for gapped Dirac fermions in energy-momentum space,
\begin{gather}\label{appendix_free_fermion_propagator}
    G_0(k) = \frac{\slashed{k} + \Delta}{k^2 - \Delta^2}.
\end{gather}
Therefore we find that at zero external 4-momenta,
\begin{gather}\label{vector_polarization}
    \Pi_V^{\mu\nu}(q) = -i\int_k \frac{\textrm{tr}[(\slashed{k} + \Delta)\gamma^{\mu}(\slashed{k} + \Delta)\gamma^{\nu}]}{(k^2 - \Delta^2)^2} + O(q^2) = O(q^2).
\end{gather}
The latter equality comes once dimensional regularization, which upholds gauge covariance, is applied \cite{bellac1992quantum}. Now we calculate the fully axial polarization, i.e. the diagram in Figure \ref{fig:bubble_diagram} at zero external 4-momentum,
\begin{gather}\nonumber
    \Pi^{\mu\nu}(q) = -i\int_k \frac{\textrm{tr}[(\slashed{k} + \Delta)\gamma^{\mu}\gamma^5(\slashed{k} + \Delta)\gamma^{\nu}\gamma^5]}{(k^2 - \Delta^2)^2} +O(q^2) \\ \label{pseudovector_polarization}
    =
    -\Pi_V^{\mu\nu}(0) - 4i\int_k \frac{2\Delta^2\eta^{\mu\nu}}{(k^2 - \Delta^2)^2} + O(q^2) = \eta^{\mu\nu}\frac{\Delta^2}{2\pi^2}\ln\left(\frac{\Lambda^2}{\Delta^2 }\right)+ O(q^2).
\end{gather}
The axion decay constant $f_\theta$ can be extracted by comparing to (\ref{main_text_polarization}),
$$
f_\theta^2 = \frac{\Delta^2}{8\pi^2}\ln\left(\frac{\Lambda^2}{\Delta^2 }\right).
$$
The velocities $v_i$ are just equal to the Fermi velocity, due to the isotropy of the model.

\subsection{Finite Magnetic Field}\label{appendix:bubble_finite_B}
In the presence of a magnetic field $\mathbf{B} = B\hat{\mathbf{z}}$, the  fermionic spectrum is given by the relativistic Landau Levels. The propagator   $G_B(x,x') = e^{i\varphi(x,x')}\tilde{G}_B(x-x')$ is given by the product of a phase and a translation-invariant propagator. The latter can be conveniently parametrized in the Schwinger (imaginary) time propagator formalism \cite{Schwinger1951OnGauge,MIRANSKY20151},
\begin{gather}
    \tilde{G}_B(k) = \int_0^\infty ds  e^{-s(\Delta^2 + k_\parallel^2 + k_\perp^2\frac{\tanh(eBs)}{eBs})}\left\{\left[\Delta - \slashed{k}_\parallel\right]\left[1 - i\gamma^1\gamma^2 \tanh(eBs)\right] - \slashed{k}_\perp[1-\tanh^2(eBs)]\right\}
\end{gather}
where the Euclidean four momentum $k = (k^0,k^1,k^2,k^3)$ can be subdivide in parallel $k_\parallel = (k^0,k^3)$ and perpendicular components $k_\perp = (k^1,k^2)$.
For the bubble diagram in Fig. \ref{fig:bubble_diagram}, the phase part drops out and only the translation-invariant part is needed. It will be necessary to calculate the following traces
\begin{gather}\nonumber
    \textrm{tr}\left[\gamma^\mu_\parallel \gamma^5 \tilde{G}_B(k) \gamma^\nu_\parallel \gamma^5 \tilde{G}_B(k)\right] = 
    -\int ds_1 ds_2 e^{-\sum_i s_i(\Delta^2 + p_\parallel^2 + p_\perp^2\frac{\tanh(eBs_i)}{eBs_i})} \\\left\{4[-\Delta^2\delta^{\mu\nu}_\parallel + p_\parallel^2\delta^{\mu\nu}_\parallel - 2p_\parallel^\mu p_\parallel^\nu][1 + \tanh(eBs_1)\tanh(eBs_2)] + p_\perp^2\delta^{\mu\nu}_\parallel[1 - \tanh^2(eBs_1)][1 - \tanh^2(eBs_2)] \right\}, \\
    \nonumber
    \textrm{tr}\left[\gamma^\mu_\perp \gamma^5 \tilde{G}_B(k) \gamma^\nu_\perp \gamma^5 \tilde{G}_B(k)\right] = 
    -\int ds_1 ds_2 e^{-\sum_i s_i(\Delta^2 + p_\parallel^2 + p_\perp^2\frac{\tanh(eBs_i)}{eBs_i})} \\\left\{4\delta^{\mu\nu}_\perp[-\Delta^2 + p_\parallel^2][1 - \tanh(eBs_1)\tanh(eBs_2)] + [p_\perp^2\delta^{\mu\nu}_\perp - 2p_\perp^\mu p_\perp^\nu][1 - \tanh^2(eBs_1)][1 - \tanh^2(eBs_2)] \right\}. 
\end{gather}
Similarly as for momentum, $\gamma_\parallel^\mu =(\gamma^0,\gamma^3$ and $\gamma_\perp^\mu =(\gamma^1,\gamma^2)$. As in the $eB = 0$ case (\ref{pseudovector_polarization}), the axial polarization can be expressed as the purely vector polarization plus a correction,
\begin{gather}\nonumber
    \Pi^{\mu\nu}_\parallel(0) = \int\frac{d^4k}{(2\pi)^4} \textrm{tr}\left[\gamma^\mu_\parallel \gamma^5  G_B(k) \gamma^\nu_\parallel \gamma^5  G_B(k)\right] 
    =  -\int\frac{d^4k}{(2\pi)^4}\textrm{tr}\left[\gamma^\mu_\parallel \tilde{G}_B(k) \gamma^\nu_\parallel \tilde{G}_B(k)\right] +2\delta^{\mu\nu}_\parallel\frac{eB}{(2\pi)^2}\int ds e^{-s\Delta^2}\frac{\Delta^2}{\tanh(eBs)} \\
    = 2\delta^{\mu\nu}_\parallel\frac{eB}{(2\pi)^2}\int ds e^{-s\Delta^2}\frac{\Delta^2}{\tanh(eBs)} = 2\delta^{\mu\nu}_\parallel\frac{\Delta^2}{(2\pi)^2}\left[\frac{eB}{\Delta^2} + \ln\left(\frac{\Delta^2}{2eB}\right) - \psi\left(1 + \frac{\Delta^2}{2eB}\right) + \Gamma\left(0,\frac{\Delta^2}{\Lambda^2}\right)\right]\\
    \nonumber
    \Pi^{\mu\nu}_\perp(0) = \int\frac{d^4k}{(2\pi)^4} \textrm{tr}\left[\gamma^\mu_\perp \gamma^5 \tilde{G}_B(k) \gamma^\nu_\perp \gamma^5 \tilde{G}_B(k)\right] 
   = 
   \int\frac{d^4k}{(2\pi)^4}\textrm{tr}\left[\gamma^\mu_\perp \tilde{G}_B(k) \gamma^\nu_\perp \tilde{G}_B(k)\right] +2\delta^{\mu\nu}_\perp\frac{eB}{(2\pi)^2}\int ds e^{-s\Delta^2}\frac{\Delta^2}{eBs} \\
   = +2\delta^{\mu\nu}_\perp\frac{eB}{(2\pi)^2}\int ds e^{-s\Delta^2}\frac{\Delta^2}{eBs} = \delta^{\mu\nu}_\perp\frac{2\Delta^2}{(2\pi)^2}\Gamma\left(0,\frac{\Delta^2}{\Lambda^2}\right) \approx 
   \delta^{\mu\nu}_\perp\frac{2\Delta^2}{(2\pi)^2}\ln\left(\frac{\Lambda^2 e^{-\gamma}}{\Delta^2}\right).
\end{gather}
The axion decay constant can be extracted once again,
\begin{gather}
    f_\theta^2 = \frac{\Delta^2}{8\pi^2}\left[\frac{eB}{\Delta^2} + \ln\left(\frac{\Delta^2}{2eB}\right) - \psi\left(1 + \frac{\Delta^2}{2eB}\right) + \Gamma\left(0,\frac{\Delta^2}{\Lambda^2}\right)\right].
\end{gather}
The velocity in the direction parallel to $\mathbf{B}$, i.e. $v_3$ in this case, remains the same as the Fermi velocity. The velocity on the transverse plane is however now given by $v_1^2 = v_2^2 = \frac{\Pi^{11}_\perp(0)}{\Pi^{00}_\parallel(0)}$.

\section{Fermionic Triangle Diagrams}\label{appendix_triangle_diagrams}
Here we calculate the contribution from Figure \ref{fig:triangle_diagrams} to lowest order in the external momenta $k_1$, $k_2$ and $q = k_1 + k_2$. To have more compact expressions, we will be using a notation where the momentum labels substitute the greek indices when contracted, e.g. $\epsilon^{\mu\nu\rho k} \equiv \epsilon^{\mu\nu\rho \sigma} k_\sigma$ or  $\epsilon^{\mu\nu p k_1 + k_2} \equiv \epsilon^{\mu\nu\rho \sigma} p_\rho (k_1 + k_2)_\sigma$. To do so, it is first convenient to study the related but more often discussed diagrams related to the chiral anomaly, which are the same as those of Figure \ref{fig:triangle_diagrams} but with the vertex replacement $\gamma^{\nu}\gamma^5, \gamma^{\rho}\gamma^5,  \rightarrow \gamma^{\rho}, \gamma^{\rho}$. The contribution of these diagrams, which we label as $I^{\mu\nu\rho}$, reads
\begin{gather}\nonumber
    I^{\mu\nu\rho} = i\int_l \frac{\textrm{tr}\left[(\slashed{l} +\Delta) \gamma^5\gamma^\mu(\slashed{l} - \slashed{k}_1 - \slashed{k}_2 +\Delta) \gamma^\rho(\slashed{l} - \slashed{k}_1 +\Delta) \gamma^\nu\right]}{[l^2-\Delta^2][(l-k_1-k_2)^2-\Delta^2][(l-k_1)^2-\Delta^2]}+\frac{\textrm{tr}\left[(\slashed{l} +\Delta) \gamma^5\gamma^\mu(\slashed{l} - \slashed{k}_1 - \slashed{k}_2 +\Delta) \gamma^\nu(\slashed{l} - \slashed{k}_2 +\Delta) \gamma^\rho\right]}{[l^2-\Delta^2][(l-k_1-k_2)^2-\Delta^2][(l-k_2)^2-\Delta^2]} \\
    \nonumber
    = -4\int_l \frac{l^2(2\epsilon^{\mu\nu\rho k_1} + \epsilon^{\mu\nu\rho k_2} - \epsilon^{\mu\nu\rho l}) + 2l^\nu\epsilon^{\mu\rho l k_1 + k_2} - 2l^\mu \epsilon^{\nu \rho l k_1} - \Delta^2(\epsilon^{\mu\nu\rho k_2} - \epsilon^{\mu\nu\rho l}) }{[l^2-\Delta^2][(l-k_1-k_2)^2-\Delta^2][(l-k_1)^2-\Delta^2]} \\
    \nonumber
    + 4\int_l \frac{l^2(2\epsilon^{\mu\nu\rho k_2} + \epsilon^{\mu\nu\rho k_1} - \epsilon^{\mu\nu\rho l}) - 2l^\rho\epsilon^{\mu\nu l k_1 + k_2} - 2l^\mu \epsilon^{\nu \rho l k_2} - \Delta^2(\epsilon^{\mu\nu\rho k_1} - \epsilon^{\mu\nu\rho l}) }{[l^2-\Delta^2][(l-k_1-k_2)^2-\Delta^2][(l-k_2)^2-\Delta^2]} + O(k^2)
    \\
    \nonumber
    = 4\int_l \frac{(l^2+\Delta^2)\epsilon^{\mu\nu\rho k_2 - k_1} - 2l^\rho \epsilon^{\mu\nu l k_1 + k_2} - 2l^\nu \epsilon^{\mu\rho l k_1 + k_2}- 2l^\mu \epsilon^{\nu\rho l k_2 - k_1} - 2l\cdot(k_2-k_1)\epsilon^{\mu\nu\rho l}}{[l^2 - \Delta^2]^3} +O(k^2).
\end{gather}
When contracting with $k_{1 \nu}$ or $(k_1 + k_2)_ \mu$, one obtains the expected result through the standard method (in the small $k$ limit). Furthermore, using $\int_l l^\mu l^\nu f(l^2) = \frac{1}{d} \int_l l^2 f(l^2)$,
\begin{gather}\label{appendix_anomaly_integral}
    I^{\mu\nu\rho} = 4\int_l \frac{\Delta^2 \epsilon^{\mu\nu \rho k_2 - k_1} }{[l^2 - \Delta^2]^3}.
\end{gather}
Now we consider the case where the first integral (corresponding to the first diagram in Figure \ref{fig:triangle_diagrams}) is shifted $l \rightarrow l+a$ and the second one is shifted $l\rightarrow l+b$. We thus obtain
\begin{gather}\nonumber
    I^{\mu\nu\rho}(a,b) 
    = I^{\mu\nu\rho} + 4\int_l \frac{(l^2-\Delta^2)\epsilon^{\mu\nu\rho a - b} + 2(a-b)\cdot l \epsilon^{\mu\nu\rho l} - 6(a-b)\cdot l \epsilon^{\mu\nu\rho l}}{[l^2-\Delta^2]^3} + O(k^2, a^2,b^2)
\end{gather}
which as before, becomes
\begin{gather}\label{appendix_anomaly_integral_after_shift}
     I^{\mu\nu\rho}(a,b) = I^{\mu\nu\rho} - 4\int_l \frac{\Delta^2\epsilon^{\mu\nu\rho a - b}}{[l^2-\Delta^2]^3}.
\end{gather}
Defining $a = \alpha(k_1 + k_2) + \beta(k_1 - k_2)$ and $b = \alpha(k_1 + k_2) + \beta(k_2 - k_1)$ (notice $a$ and $b$ satisfy momentum swapping $k_1 \leftrightarrow k_2$), one obtains the following conservation laws
\begin{gather}\label{appendix_vector_conservation}
    U(1)_V : \quad k_{1\nu}I^{\mu\nu\rho}(a,b) = 4\int_l \frac{\Delta^2}{[l^2 - \Delta^2]^3}\left[-1 - 2\beta\right]\epsilon^{\mu\rho k_1 k _2} , \\
    \label{appendix_axial_conservation}
    U(1)_A : \quad (k_{1} + k_2)_\mu I^{\mu\nu\rho}(a,b) = 4\int_l \frac{\Delta^2}{[l^2 - \Delta^2]^3}\left[2 + 4\beta\right]\epsilon^{\nu\rho k_1 k _2} .
\end{gather}

We see then that to enforce $U(1)_V$ gauge invariance, needed for a theory with electric charge conservation, one must enforce $\beta = -1/2$ \cite{Schwartz_2013}.
But this implies that $ I^{\mu\nu\rho}(a,b) = 0$, so that even the axial anomaly contribution (\ref{appendix_axial_conservation}) is vanishing $(k_1+k_2)_\mu I^{\mu\nu\rho}(a,b) = 0$ at this order \cite{Yu2021Dynamical}. As a further check, one can divide the $U(1)_A$ expression into its purely anomalous and its matter part,
\begin{gather}
    \quad (k_{1} + k_2)_\mu I^{\mu\nu\rho}(a,b) = 4\int_l \frac{\Delta^2}{[l^2 - \Delta^2]^3}\left[-2 + 4\beta\right]\epsilon^{\nu\rho k_1 k _2} + 16\int_l \frac{\Delta^2}{[l^2 - \Delta^2]^3}\epsilon^{\nu\rho k_1 k _2},
\end{gather}
where the latter integral is equal to
\begin{gather}
    16\int_l \frac{\Delta^2}{[l^2 - \Delta^2]^3}\epsilon^{\nu\rho k_1 k _2}  = -2i\Delta\int_l \textrm{tr}\left[\frac{1}{\slashed{l} - \Delta} \gamma^5\frac{1}{\slashed{l} - \slashed{k}_1 -\slashed{k_2} - \Delta} \gamma^\rho\frac{1}{\slashed{l} - \slashed{k}_1  - \Delta} \gamma^\nu \right] + (k_1, \nu \leftrightarrow k_2, \rho),
\end{gather}
which is the expected non-anomalous term which exists simply due to the massive nature of the Dirac fermions.
Now we turn to calculating the purely axial contribution, which we denote as $\Gamma^{\mu\nu\rho}$, from  the diagrams in Figure \ref{fig:triangle_diagrams},
\begin{gather}\nonumber
    \Gamma^{\mu\nu\rho} = i\int_l \frac{\textrm{tr}\left[(\slashed{l} +m) \gamma^5\gamma^\mu(\slashed{l} - \slashed{k}_1 - \slashed{k}_2 +m) \gamma^5\gamma^\rho(\slashed{l} - \slashed{k}_1 +m) \gamma^5\gamma^\nu\right]}{[l^2-\Delta^2][(l-k_1-k_2)^2-\Delta^2][(l-k_1)^2-\Delta^2]}\\
    -\frac{\textrm{tr}\left[(\slashed{l} +m) \gamma^5\gamma^\mu(\slashed{l} - \slashed{k}_1 - \slashed{k}_2 +m) \gamma^5\gamma^\nu(\slashed{l} - \slashed{k}_2 +m) \gamma^5\gamma^\rho\right]}{[l^2-\Delta^2][(l-k_1-k_2)^2-\Delta^2][(l-k_2)^2-\Delta^2]} \\
    \nonumber
    = -4\int_l \frac{(l^2+\Delta^2)(2\epsilon^{\mu\nu\rho k_1} + \epsilon^{\mu\nu\rho k_2} - \epsilon^{\mu\nu\rho l}) + 2l^\nu\epsilon^{\mu\rho l k_1 + k_2} - 2l^\mu \epsilon^{\nu \rho l k_1} - 2\Delta^2\epsilon^{\mu\nu\rho l} }{[l^2-\Delta^2][(l-k_1-k_2)^2-\Delta^2][(l-k_1)^2-\Delta^2]} \\
    \nonumber
    + 4\int_l \frac{(l^2+\Delta^2)(2\epsilon^{\mu\nu\rho k_2} + \epsilon^{\mu\nu\rho k_1} - \epsilon^{\mu\nu\rho l}) - 2l^\rho\epsilon^{\mu\nu l k_1 + k_2} - 2l^\mu \epsilon^{\nu \rho l k_2} - 2\Delta^2\epsilon^{\mu\nu\rho l} }{[l^2-\Delta^2][(l-k_1-k_2)^2-\Delta^2][(l-k_2)^2-\Delta^2]} + O(k^2)
    \\
    \nonumber
    = -4\int_l -\frac{(l^2+\Delta^2)\epsilon^{\mu\nu\rho k_2 - k_1} - 2l^\rho \epsilon^{\mu\nu l k_1 + k_2} - 2l^\nu \epsilon^{\mu\rho l k_1 + k_2}- 2l^\mu \epsilon^{\nu\rho l k_2 - k_1} - 2l\cdot(k_2-k_1)\epsilon^{\mu\nu\rho l}}{[l^2 - \Delta^2]^3} \\ + \frac{4\Delta^2 2l\cdot(k_2-k_1)\epsilon^{\mu\nu\rho l}}{[l^2 - \Delta^2]^4} +O(k^2) =  I^{\mu\nu\rho} -  4\int_l \frac{4\Delta^2 2l\cdot(k_2-k_1)\epsilon^{\mu\nu\rho l}}{[l^2 - \Delta^2]^4} +O(k^2) .
\end{gather}
Notice that in the $\Delta\rightarrow 0$ limit, $\Gamma^{\mu\nu\rho} \rightarrow I^{\mu\nu\rho}$, as expected. Introducing momentum translations as in the standard chiral anomaly triangle diagram (\ref{appendix_anomaly_integral_after_shift}),
\begin{gather}
    \Gamma^{\mu\nu\rho}(a,b) =  I^{\mu\nu\rho}(a,b) - 4\int_l \frac{4\Delta^2 2l\cdot(k_2-k_1 + 3a - 3b)\epsilon^{\mu\nu\rho l}}{[l^2 - \Delta^2]^4} +O(a^2, b^2, k^2).
\end{gather}
Choosing $\beta = -1/2$ to maintain $U(1)_V$ gauge invariance as before, then $I^{\mu\nu\rho}(a,b) = 0$ as discussed above and
\begin{gather}
    \Gamma^{\mu\nu\rho}(a,b) =  -4\int_l \frac{4\Delta^2 2l\cdot 4(k_2-k_1)\epsilon^{\mu\nu\rho l}}{[l^2 - \Delta^2]^4} +O(a^2, b^2, k^2) = i\frac{2\pi^2}{(2\pi)^4} \frac{4^2}{6}\epsilon^{\mu\nu\rho k_2-k_1}.
\end{gather}
This will thus be the value of the vertex connecting 3 axial gauge fields $A^\mu_5$.

Finally, we consider two further triangle diagrams, this time composed purely by vectorial vertices. These diagrams are as those of Figure \ref{fig:triangle_diagrams} but removing the $\gamma^5$ matrix from the vertices. Their integral reads
\begin{gather}\nonumber
    K^{\mu\nu\rho} = i\int_l \frac{\textrm{tr}\left[(\slashed{l} +\Delta) \gamma^\mu(\slashed{l} - \slashed{k}_1 - \slashed{k}_2 +\Delta) \gamma^\rho(\slashed{l} - \slashed{k}_1 +\Delta) \gamma^\nu\right]}{[l^2-\Delta^2][(l-k_1-k_2)^2-\Delta^2][(l-k_1)^2-\Delta^2]}+\frac{\textrm{tr}\left[(\slashed{l} +m) \gamma^\mu(\slashed{l} - \slashed{k}_1 - \slashed{k}_2 +m) \gamma^\nu(\slashed{l} - \slashed{k}_2 +m) \gamma^\rho\right]}{[l^2-\Delta^2]^3[(l-k_1-k_2)^2-\Delta^2][(l-k_2)^2-\Delta^2]} \\
    \nonumber
    = -4i\int_l \frac{\eta^{\mu\rho}((k_1+k_2)^\nu(\Delta^2-l^2) -2l^\nu l\cdot(k_1+k_2)) + 4l^\mu l^\rho(k_1 +k_2)^\nu + \textrm{sym. permutations of indices}}{[l^2-\Delta^2]^3} \\ \nonumber
     -4i\int_l \frac{[(\eta^{\mu\rho}l^\nu + \textrm{sym. permutations of indices})(l^2-\Delta^2) -4l^\mu l^\nu l^\rho]6l\cdot(k_1+k_2)}{[l^2-\Delta^2]^4} + O(k^2) \\ \nonumber
     = -4i\int_l \frac{\Delta^2\eta^{\mu\rho}(k_1+k_2)^\nu + \textrm{sym. permutations of indices}}{[l^2-\Delta^2]^3} - \frac{\Delta^2l^2\eta^{\mu\rho}(k_1+k_2)^\nu + \textrm{sym. permutations of indices}}{[l^2-\Delta^2]^4} + O(k^2)\\
     = -4i\int_l \frac{\Delta^4\eta^{\mu\rho}(k_1+k_2)^\nu + \textrm{sym. permutations of indices}}{[l^2-\Delta^2]^4} + O(k^2).
\end{gather}
The "+ sym. perturbation of indices" stands for adding the remaining permutations of the indices $\mu,\nu,\rho$. Note that the third line comes from the expansion of the denominator to first order in the external momenta. The second to last line is achieved by using $\int_l l^\mu l^\nu f(l^2) = \frac{1}{2}\int_l l^2 f(l^2)$. Now, as in the case of the anomalous diagrams, we consider the case where the first integral is translated by $l \rightarrow l + a$ and the second one is translated by $l \rightarrow l + b$. The new integral is equal to 
\begin{gather}\nonumber
    K^{\mu\nu\rho}(a,b) = K^{\mu\nu\rho}
    - 4i\int_l \frac{\eta^{\mu\rho}(a^\nu(\Delta^2-l^2) -2l^\nu l\cdot a) - 4l^\mu l^\rho a^\nu + \textrm{sym. permutations of indices} + (a \leftrightarrow b)}{[l^2-\Delta^2]^3} \\ \nonumber
     +4i\int_l \frac{[(\eta^{\mu\rho}l^\nu + \textrm{sym. permutations of indices})(l^2-\Delta^2) -4l^\mu l^\nu l^\rho]6l\cdot(a+b)}{[l^2-\Delta^2]^4} + O(k^2,a^2,b^2) \\
     =  K^{\mu\nu\rho} +4i\int_l \frac{\Delta^4\eta^{\mu\rho}(a+b)^\nu + \textrm{sym. permutations of indices}}{[l^2-\Delta^2]^4} + O(k^2,a^2,b^2).
\end{gather}
As previously, $a = \alpha(k_1 + k_2) + \beta(k_1 - k_2)$ and $b = \alpha(k_1 + k_2) + \beta(k_2 - k_1)$ (again, notice $a$ and $b$ satisfy momentum swapping). Therefore
\begin{gather}\nonumber
    K^{\mu\nu\rho}(a,b) = K^{\mu\nu\rho} +4i(2\alpha)\int_l \frac{\Delta^4\eta^{\mu\rho}(k_1+k_2)^\nu + \textrm{sym. permutations of indices}}{[l^2-\Delta^2]^4} + O(k^2,a^2,b^2).
\end{gather}
In contrast to the anomaly diagram, this one depends only on $\alpha$ and not $\beta$. Thus our previous choice of $\beta = -1/2$ is of no consequence and we have the freedom to choose $\alpha =  1/2$ so that $K^{\mu\nu\rho}(a,b) = O(k_1^2,k_2^2)$. Therefore $U(1)_V$, i.e. charge conservation is preserved, i.e. contractions with $(k_1 +k_2)_\mu, k_{1\nu}$ and $k_{2\rho}$ (i.e. the 4-momenta of the vector fields) will trivially be zero, as was the case with the anomaly triangle diagrams (\ref{appendix_vector_conservation}). Although here we have sketched a derivation to be consistent with our treatment of the other triangle diagrams, it is worth noting that this diagram should be 0 to all orders in $k$ due to Furry's theorem.

\section{Imaginary component of Boson Bubble}
The contribution of the boson bubble, represented in Figure \ref{fig:bosonic_diagrams}a, is given by
\begin{gather}\nonumber
    [\Pi_\textrm{B}]_{a\bar{a}}(q) = \frac{(3!)^2}{3^2}\int_l\Gamma^{\mu\nu\rho}(l,q-l)\Gamma^{\bar{\mu}\bar{\nu}\bar{\rho}}(-l,l-q) \\
    \label{appendix_bubble_expression}
    \cdot \mathcal{A}_{\mu a}^{5}(q)\mathcal{A}_{\bar{\mu} \bar{a}}^{5}(-q)
    \mathcal{A}_{\nu b}^{5}(-l)\mathcal{A}_{\bar{\nu} \bar{b}}^{5}(l)\mathcal{A}_{\rho c}^{5}(l-q)\mathcal{A}_{\bar{\rho} \bar{c}}^{5}(q-l) 
     D_{b\bar{b}}(l)\cdot D_{\bar{c}c}(l-q).
\end{gather}
Greek indices indicate spacetime coordinates, e.g. $\mu=0,1,3,4$, whereas latin indices label the undiagonalized bosonic fields that appear in the effective action (\ref{effective_action}), i.e. $U_a$ where $a = 1,2,3,4$. The integral over $l = (i\omega_n, \mathbf{l})$ is a shorthand for a Matsubara summation over bosonic frequencies $i\omega_n = 2n\pi/T$ and an integral over momentum space
\begin{gather}
    \int_l \equiv -\frac{1}{T}\sum_{i\omega_n}\int\frac{d^3l}{(2\pi)^3}
\end{gather}
In addition, $D_{a\bar{a}}(q)$ are the boson propagators in the imaginary time formalism, evaluated at $q=(iq^0,\mathbf{q})$, are given by
\begin{gather}
    D_{a\bar{a}}(q) = \sum_m \Theta(\Lambda_m/c_m - |\mathbf{q}|)\frac{(|m,\mathbf{q}\rangle\langle m,\mathbf{q}|)_{a\bar{a}}}{(iq_0)^2 - c_m^2\mathbf{q}^2},
\end{gather}
where in this expression the index $m$ runs over the free boson modes, i.e. the longitudinal and transverse phonons, as well as the axion. The $|m,\mathbf{l}\rangle$ vector is the polarization of the mode $m$ in the bosonic field space, $c_m$ is its velocity, and $\Lambda_m$ is its energy cutoff (and hence $\Lambda_m/c_m$ the momentum cutoff). We assume that for phonons, i.e. $m =$ L, T1, T2, $\Lambda_m = T_\textrm{D} = 150 $K = 12.8 meV, where $T_\textrm{D}$ is a Debye temperature \cite{smontara2002anisotropy}. On the other hand, for the axion the cutoff is $\Lambda_\textrm{ax} = \Delta = 260/2 = 130 $ meV. This value is extracted from the value of the gap of the charge density wave in
(TaSe$_4$)I \cite{gooth2019axionic}, which is 260 meV. Note that despite the energy cutoff for axions being much greater than for axions, the momentum cutoff for the latter is actually much smaller that the former's due to the fact that $c_{\textrm{L,T}}/v_f \sim 10^{-2}$. In the following we outline some analytical manipulations used to ease the numerical evaluation of (\ref{appendix_bubble_expression}).

\subsection{Matsubara Summation}
For the evaluation of (\ref{appendix_bubble_expression}), the evaluation of the following type of expressions involving the denominators of the boson propagators will be relevant
\begin{gather}\nonumber
    I_{ab} = -\beta\sum_{i\omega_n} \frac{F(i\omega_n)}{[(i\omega_n)^2 - \omega_{a\mathbf{l}}^2][(i\omega_n - \omega)^2 - \omega_{b\mathbf{l}-\mathbf{q}}^2]}
    = \frac{1}{2\omega_{a\mathbf{l}}} \left[\frac{F(-\omega_{a\mathbf{l}})}{(\omega_{a\mathbf{l}}+\omega)^2 - \omega_{b\mathbf{l}-\mathbf{q}}^2}\right] + \frac{1}{2\omega_{b\mathbf{l}-\mathbf{q}}} \left[\frac{F(\omega-\omega_{b\mathbf{l}-\mathbf{q}})}{(\omega_{b\mathbf{l}-\mathbf{q}}-\omega)^2 - \omega_{a\mathbf{l}}^2}\right]  \\
    \label{matsubara}
    +\frac{b_0(\omega_{a\mathbf{l}})}{2\omega_{a\mathbf{l}}} \left[\frac{F(\omega_{a\mathbf{l}})}{(\omega_{a\mathbf{l}}-\omega)^2 - \omega_{b\mathbf{l}-\mathbf{q}}^2}+\frac{F(-\omega_{a\mathbf{l}})}{(\omega_{a\mathbf{l}}+\omega)^2 - \omega_{b\mathbf{l}-\mathbf{q}}^2}\right] 
    +\frac{b_0(\omega_{b\mathbf{l}-\mathbf{q}})}{2\omega_{b\mathbf{l}-\mathbf{q}}} \left[\frac{F(\omega + \omega_{b\mathbf{l}-\mathbf{q}})}{(\omega_{b\mathbf{l}-\mathbf{q}}+\omega)^2 - \omega_{a\mathbf{l}}^2}+\frac{F(\omega -\omega_{b\mathbf{l}-\mathbf{q}})}{(\omega_{b\mathbf{l}-\mathbf{q}}-\omega)^2 - \omega_{a\mathbf{l}}^2} \right] .
\end{gather}
In this expression, $F(i\omega_n)$ is some numerator that, importantly, depends on the Matsubara frequency and $a,b$ are indices indicating phononic or axion modes.

\subsection{Imaginary Component}
The numerator $F(i\omega_n)$, once an analytic continuation for the external frequency back to real space (it should be understood that $\omega \equiv \omega + i\varepsilon$) has been performed, has no imaginary component, which will hence be extracted exclusively from the denominator. We use the Sokhotski–Plemelj theorem,
\begin{equation}
    \frac{1}{X + i\epsilon} = -i\pi\delta(X) + \mathcal{P} \frac{1}{X}.
\end{equation}
This implies that the imaginary part of (\ref{matsubara}) becomes
\begin{gather}\nonumber
    \textrm{Im} I_{ab} = \frac{-\pi}{4\omega_a \omega_b} \sum_{s_1 = \pm}\left\{\left[b_0(s_1\omega_a)F(s_1\omega_a)\sum_{s_2=\pm} s_1s_2\delta(\omega - s_1\omega_a - s_2\omega_b)\right] +\right. \\
    \nonumber
    \left.+ \left[b_0(s_1\omega_b)F(\omega + s_1\omega_b)\sum_{s_2=\pm} s_1s_2\delta(\omega - s_2\omega_a + s_1\omega_b)\right]\right\}.
\end{gather}
Rearranging, result can be recast as
\begin{gather}\label{bubble_imaginary_component_omega}
    \textrm{Im} I_{ab} = -\frac{\pi}{4\omega_a\omega_b}\sum_{s_a,s_b = \pm}s_a s_b\delta(\omega - s_a\omega_a - s_b\omega_b)F(s_a\omega_a)\left[b_0(s_a\omega_a) - b_0(-s_b\omega_b)\right]
\end{gather}
Using that $\omega_a = c_al$ and $\omega_b = c_b|\mathbf{l} - \mathbf{q}|$, we find that the Dirac Deltas can be reduced to one for a cosine of the angle between $\mathbf{q}$ and $\mathbf{l}$,
\begin{gather}\label{dirac_delta_cosine}
    \delta(\omega - s_a\omega_a -s_b\omega_b) = \frac{\omega_b}{c_b^2 ql}\delta(\cos\theta_{lq} - \cos\theta_{s_a s_b})\Theta(s_b(\omega - s_a\omega_a)),
\end{gather}
where
\begin{gather}
    \cos\theta_{s_a s_b} = \frac{l^2+q^2 - \omega_b^2/c_b^2}{2lq}
\end{gather}
and $\omega_b = s_b(\omega - s_a \omega_a)$.This implies that (\ref{bubble_imaginary_component_omega}) becomes
\begin{gather}\label{bubble_imaginary_component_omega2}
    \textrm{Im} I_{ab} = -\frac{\pi}{4\omega_a c_b^2 ql}\sum_{s_a,s_b = \pm}s_a s_b\delta(\cos\theta - \cos\theta_{s_a s_b})F(s_a\omega_a)\left[b_0(s_a\omega_a) - b_0(-s_b\omega_b)\right]\Theta(s_b(\omega - s_a\omega_a)).
\end{gather}

\end{widetext}

\bibliography{apssamp}

\end{document}